\documentclass{aa}  
\usepackage{graphicx}
\usepackage{txfonts}
\usepackage{gensymb}

\begin{document}

   \title{Reconstructing solar magnetic fields from historical observations
\\X. Effect of magnetic field inclination and boundary structure on AIA 1600 Å emission}
   
   \author{Ismo Tähtinen
          \inst{1}
            \and
          Alexei A. Pevtsov\inst{2}
          \and
          Timo Asikainen\inst{1}
          \and
          Kalevi Mursula\inst{1,2}
          }

   \institute{Space Physics and Astronomy Research Unit, University of Oulu,
               POB 8000, FI-90014, Oulu, Finland\\
              \email{ismo.tahtinen@oulu.fi}
            \and
             National Solar Observatory, Boulder, CO 80303, USA \\
             }


 
  \abstract
   {The relation between the intensity of chromospheric emissions and the photospheric magnetic field strength has been examined in several studies, but the effect of the magnetic field inclination on chromospheric emissions remains almost unexplored.
   }
   {We study how the inclination of the photospheric magnetic field, as measured by the full 3D magnetic vector from the Helioseismic and Magnetic Imager (HMI), affects the relationship between the magnetic field strength and the far-ultraviolet emission at around 1600~Å observed by the Atmospheric Imaging Assembly (AIA). 
   We also study how these parameters change spatially close to the active region perimeter.
   }
   {We analyzed the mutual dependence of 1168 co-temporal AIA and HMI observations from 2014 to 2017.
   We focused on magnetically active regions outside sunspots (e.g., plages and network) close to the solar disk center.
   We studied how the AIA and HMI parameters change with distance from the active region perimeter.
   }
   {The AIA 1600 emission typically decreases with increasing (more horizontal) inclination.
   For all inclinations, AIA~1600 emission increases with increasing magnetic field strength until saturating at some peak intensity, which depends on the cosine of the inclination, with horizontal regions saturating at lower intensities.
   In addition, we find that activity clusters have a narrow boundary (<~2 arcseconds) in which the AIA~1600 intensity, magnetic field strength, and inclination distributions and relations differ significantly from those in the inner layers.
   }
   {This study demonstrates the significant effect that magnetic field inclination and activity cluster border regions have on chromospheric emissions.
   Although the observed effects are likely reduced in low-resolution observations where different regions are averaged together, a detailed study is needed to examine the emission--magnetic field relation at different resolutions.
   }

               
    \titlerunning{Effect of magnetic field inclination on SDO/AIA 1600 emission}
   \maketitle
%

\section{Introduction}\label{s:introduction} 
Since the first sunspot magnetic field strength measurements at the Mount Wilson Observatory, it has been known that magnetic fields can affect solar radiation \citep{Hale1908}.
Following the invention of an electronic magnetograph \citep{Babcock1953}, it was established that magnetic fields could not only decrease radiation, like in sunspots, but also enhance solar emission.
Plages visible in chromospheric emissions such as \rm{Ca II K} were found to correspond spatially with areas of strong magnetic fields in magnetographic observations \citep{Babcock.Babcock1955,Leighton1959,Stepanov.Petrova1959}.
Since then, a strong correlation between the intensity of chromospheric emissions and the (unsigned) magnetic field strength has been established in numerous studies \citep{Skumanich1975,Schrijver_1989,HarveyWhite1999,Ortiz_2005,Rezaei_2007,Loukitcheva2009,Pevtsov.etal2016,Kahil_2017,Kahil2019,Chatzistergos2019,Tahtinen2022}.

Studies relating chromospheric emissions to photospheric magnetic fields have been mainly based on line-of-sight (LOS) observations of the magnetic field.
Measurements of the full 3D magnetic vector, which provide information on the orientation of magnetic fields and on the true field strength, have not been studied in this context.
Thus, the possible effect of magnetic field inclination on chromospheric emissions has remained unexplored so far.

The effect of magnetic field inclination on wave propagation in the solar atmosphere is well known by now \citep[see, e.g., the review by][]{Khomenko2013}.
Waves propagating upward from the photosphere can deposit their energy higher in the solar atmosphere, contributing to chromospheric heating.
The inclination of the magnetic field directly affects the propagation of magnetoacoustic waves, for example by reducing the cutoff frequency, which is proportional to the cosine of the angle between the magnetic field line and the vertical direction \citep{Bel1977,Rajaguru2019,Yelles2023}.
Although the effect of magnetic field inclination on the properties of magnetoacoustic waves and their contribution to chromospheric heating has been extensively studied, only a few studies have directly investigated how the inclination affects solar emissions.
\citet{Kostik2016} studied the relations between Ca~II~H line core brightness and the magnetic field strength and inclination in solar faculae close to the disk center.
They find that Ca~II~H brightness increases with increasing field strength and decreases with increasing inclination.
Their study was limited to a single small region, which they observed for about 30 minutes.

In a recent study, \citet{Tahtinen2022} find evidence that, in addition to field strength, the inclination of the magnetic field also affects the intensity of the far-ultraviolet (FUV) 1600~Å emission observed by the Atmospheric Imaging Assembly (AIA).
They find that regions with comparable magnetic field strengths seem to emit less radiation when the magnetic field is highly inclined.
However, this conclusion was based on a single co-spatial and co-temporal pair of an AIA~1600 image and a Helioseismic and Magnetic Imager (HMI) magnetogram.
Here, we study 1168 simultaneous AIA~1600 images and HMI magnetograms.
This paper is organized as follows.
Section \ref{sec:data} presents the data and discusses their limitations.
Section \ref{sec:Inclination} analyzes how the AIA~1600 intensity behaves as a function of magnetic field strength and inclination. 
Section \ref{sec:boundary} analyzes how the magnetic field and AIA~1600 distributions change when moving from the perimeter of an active region to the inside.
We discuss our results in Sect. \ref{sec:discussion} and give our conclusions in Sect.~\ref{sec:conclusions}.

\section{Data}\label{sec:data}
We used magnetic field and pseudo-continuum intensity measurements of the HMI \citep[][]{Scherrer.etal2012} and FUV measurements around 1600~Å of the AIA \citep{Lemen.etal2012}, both of which are on board the Solar Dynamics Observatory \citep[SDO;][]{Pesnell.etal2012}. 
Our dataset consists of 1168 pairs of simultaneous HMI--AIA observations taken from 1~March~2014 to 9~June~2017 at 05:48 UTC once a day.
Using this particular time for all observations minimizes the variation in magnetic field strength due to the orbital motion of the SDO, which can be up to 5\% with a period of 12-hours \citep{Hoeksema2014,Smirnova.etal2013}.
We downloaded the data from the Joint Science Operations Center (JSOC\footnote{\url{http://jsoc.stanford.edu/}}) and co-registered HMI and AIA images with the Python aiapy package \citep{Barnes2020}.
Registration co-aligns images and sets them to the same spatial scale of 0.6 arcsec/px.

\subsection{HMI magnetic field}
Helioseismic and Magnetic Imager 720-second magnetic field data products are calculated every 12 minutes from filtergrams taken at a 3.75-second cadence.
Filtergrams measure six wavelengths around the Fe \rm{I} 617.3 nm absorption line in different polarization states \citep{Hoeksema2014}.
We used both vector field and LOS measurements of HMI.
The upper limit for random noise in LOS measurements is 6.3~G \citep{Liu2012}, while in vector field measurements the noise level is about 100~G \citep{Hoeksema2014}.

Inversion of the vector field includes a 180\degree{} ambiguity in the azimuthal direction of the transverse component, which is resolved in the HMI data pipeline.
The success of disambiguation depends on the strength of the transverse component of the magnetic field.
Therefore, the results are most reliable for strong field regions.
HMI pipeline classifies inverted pixels into three confidence classes and uses different disambiguation methods for the three classes.
Those regions whose transverse field strength exceeds the noise level by 50~G are considered to be high-confidence regions.
Regions within~5 pixels of high-confidence pixels are intermediate-confidence regions, and the rest of the pixels are weak-confidence regions.

For high- and intermediate-confidence pixels the HMI pipeline uses a variant of the minimum energy method \citep{Metcalf1994} called ME0 to resolve the 180\degree{} ambiguity.
The ME0 method minimizes the magnitude of field divergence and the normal component of electric current density.
ME0 works well for strong field regions, but suffers from the presence of noise.
For those pixels whose transverse field strength does not exceed the noise level by 50~G, several other methods can be used.
In intermediate-confidence regions, the minimum energy solution is smoothed using the neighboring-pixel acute-angle algorithm, which maximizes the sum of the dot product of the pixel and its neighbors.
In weak-confidence regions, three different methods are used: potential field acute-angle method, most-radial -method, and random method.
The potential field acute-angle method chooses an azimuth, which is closest to the derivative of the potential field used to approximate the field divergence.
The most radial method selects an azimuth that produces the most radial field.
The random method sets random disambiguation for the azimuth.
To avoid any systematic errors, we used the random disambiguation for the weak-confidence regions since the other two methods can produce large-scale patterns in azimuth \citep{Hoeksema2014}.

The magnetic field quantities we study in this paper are the vector magnetic field strength, $B_\mathrm{Vec}$, the strength of the LOS component, $B_\mathrm{LOS}$, the local inclination of the magnetic field, $\gamma$ (with respect to the radial direction), and the inclination with respect to the LOS direction, $\gamma_\mathrm{LOS}$.
We computed the local inclination, $\gamma,$ as the acute angle between the magnetic field vector and the vertical direction
so that $\gamma \in [0\degree,90\degree]$, where 0\degree{} corresponds to a vertical (radial) field and 90\degree{} to a horizontal field (tangent to the solar surface).
The LOS inclination $\gamma_\mathrm{LOS}$ is available from the JSOC and does not need to be computed separately.
We ignored the sign (polarity) information of $\gamma_\mathrm{LOS}$ so that, as for the local inclination, $\gamma_\mathrm{LOS} \in [0\degree,90\degree]$, where 0\degree{} corresponds to aligned with the LOS and 90\degree{} to perpendicular to the LOS.

\subsection{HMI pseudo-continuum brightness and sunspots}
We used the HMI pseudo-continuum brightness, $I_C$, which is derived from the same Fe \rm{I} 617.3 nm filtergrams as the magnetic field measurements \citep{Couvidat2012}, to remove sunspots from the data.
We excluded sunspots (both umbrae and penumbrae) from the data to ensure the results are specific to chromospheric brightenings.
We in particular tried to avoid the effect of sunspot penumbrae, whose magnetic fields are typically highly inclined.

The limb darkening has been removed from the data at the JSOC.
In these data, the pseudo-continuum brightness $I_C$~=~1 corresponds to the peak of the pseudo-continuum brightness distribution.
We used the lower threshold of $I_C$~=~0.94 to filter out sunspots from the observations.
We selected this threshold by studying how the brightness distribution behaves on spotless days.
We find that on spotless days, the pseudo-continuum brightness distribution had a lower limit around $I_C~=~0.94$.
Based on this result, we selected $I_C~=~0.94$ as a threshold, which we used to exclude the sunspot regions from the data.
This threshold is on the higher end of the range of values used to exclude sunspots from different data.
Similar values have been used, for example, by \citet{Mathew2007} and \citet{Verma2018}, but it is high compared to the 0.87 used by \citet{Yeo2014}, who also used HMI pseudo-continuum observations.
We note that some studies have applied more conservative limits by extending sunspot areas to include surrounding magnetic features \citep[e.g.,][]{Chatzistergos2019}.
We tested this by extending sunspot areas but found that removing such larger sunspot areas had virtually no effect on our results, only reduced the statistics of highly inclined fields. Therefore, we decided to keep to the current definition.

\subsection{AIA~1600}
AIA~1600 measures the FUV continuum around 1600~Å, originating mainly from the temperature minimum of the solar atmosphere \citep{Simoes2019}.
The intensity of AIA~1600 emission is closely related to photospheric magnetic field strength, and emission patterns observed over the solar disk closely map to the magnetic network and active regions seen in solar magnetograms \citep{Tahtinen2022}.

We calibrated the AIA~1600 images according to the procedure described in \citet{Tahtinen2022}, which corrects the data for instrument degradation.
This calibration produces contrast images where pixel intensities are given with respect to the intensity of the quiet Sun: intensity $I_\mathrm{AIA}~=~1$ is the average value of the quiet Sun, while $I_\mathrm{AIA}~=~2$ is twice as bright.
We refer here to AIA 1600 contrast values as AIA 1600 intensities for the sake of simplicity.

\subsection{HMI vector field problem}
The HMI vector field measurements suffer from a systematic bias \citep{Pevtsov2021,Liu2022}.
The transverse component of the magnetic field is systematically overestimated, which affects especially the weak- and moderate-strength magnetic field measurements.
The problem stems from the presence of noise and the filling factor, which in the standard HMI pipeline is set to unity.
A filling factor close to unity is valid for sunspot umbrae and penumbrae but may substantially differ from the correct value in moderate- and weak-field regions.

\begin{figure}
\resizebox{\hsize}{!}{\includegraphics{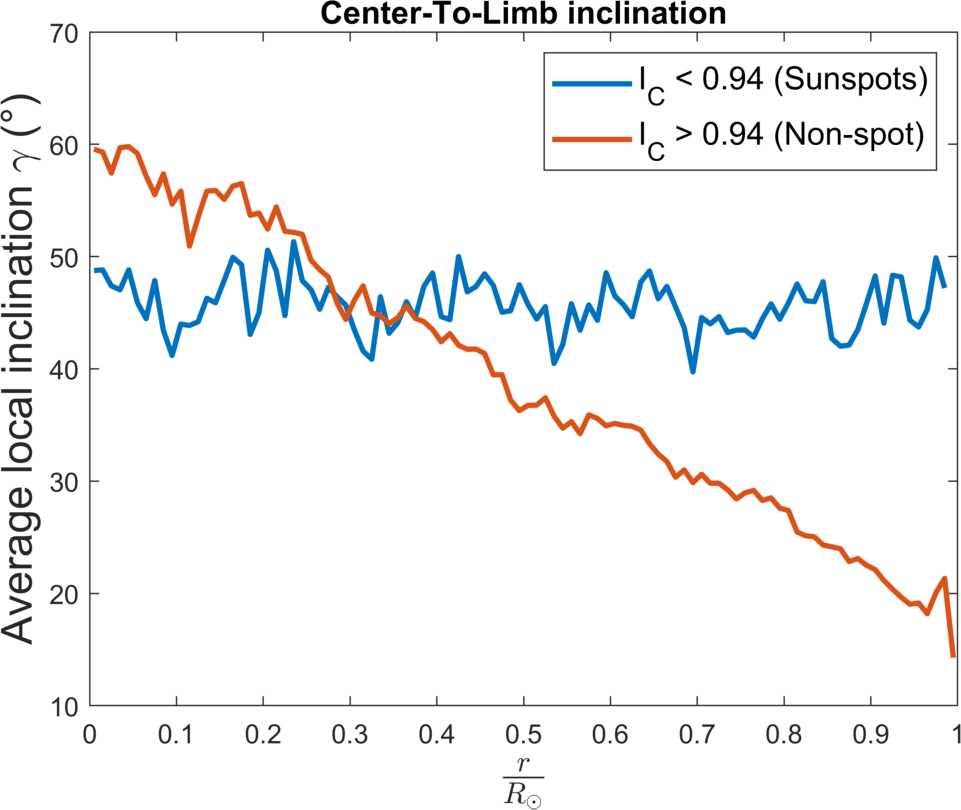}}
\caption{Center-to-limb variation in the local inclination, $\gamma$. The average local inclinations of $I_\mathrm{C} < 0.94$ (blue) and $I_\mathrm{C} > 0.94$ (orange) high-confidence pixels close to the disk equator (apparent heliographic latitude $\lambda < |\pm5.7\degree{}|$) are shown as a function of distance from the disk center.}\label{fig:CTL}
\end{figure}

The overestimation of the transverse component increases the observed radial field with increasing distance from the solar center.
This is depicted in Fig.~\ref{fig:CTL}, which shows the average local inclination $\gamma$ of the magnetic field for high-confidence sunspot ($I_\mathrm{C} < 0.94$, blue) and non-spot ($I_\mathrm{C} > 0.94$, orange) pixels close to the equator as a function of the distance from the central meridian.
Non-spot pixels mainly correspond to moderate-strength features such as plages and magnetic network.
Figure~\ref{fig:CTL} shows that the average inclination of non-spot pixels changes from about 60\degree{} at the disk center to about 20\degree{} at the solar limb while the average inclination of sunspot pixels stays roughly constant at about 45--50\degree{}.
Since the purpose of this work is to study the inclination of magnetic fields outside the sunspots, we limited our study to the region within $0.1R_\odot$ (5.7\degree{} in terms of the heliographic angle, $\theta$) of the disk center in order to mitigate the center-to-limb effect in non-spot pixels.
This choice limits the error in inclination from systematic center-to-limb variation to about 5\degree{}.
Moreover, due to this choice, the two inclination angles $\gamma$ and $\gamma_\mathrm{LOS}$ are close to each other since there is at most a 12.95\degree{} (5.7\degree{}+7.25\degree{}) difference between the radial direction and the LOS direction.
The additional 7.25\degree{} variation comes from the annual variation of the heliographic latitude of the disk center due to the Sun's rotation axis inclination with respect to the ecliptic.

\subsection{Exclusion of the quiet Sun}
Due to the poor signal-to-noise ratio and overestimation of the transverse component, the determination of the magnetic field vector in the quiet Sun is not reliable.
We excluded the quiet Sun from the dataset using two criteria.
First, we included only pixels whose LOS magnetic field strength exceeds a threshold of 50~G.
This is a typical threshold value used to separate magnetically active regions from the quiet Sun, and it also maximizes the average size of clusters consisting of adjacent pixels above a certain threshold value \citep{Tahtinen2022}.
In addition to $B_\mathrm{LOS} >$ 50~G pixels, we also included the high-confidence disambiguation pixels, that is, the pixels whose transverse component, $B_{\mathrm{Trans}}$, exceeds the vector field noise level by 50~G ($B_{\mathrm{Trans}} \gtrsim$ 150~G).
The rest of the pixels are considered to belong to the quiet Sun (or the sunspots removed earlier) and are removed from the data.

We needed to include the high-confidence disambiguation pixels to ensure that highly inclined magnetic field regions with a strong transverse field  ($\gtrsim$ 150~G) but a small $B_\mathrm{LOS}$ ($<$ 50~G) are included.
On the other hand, the high-confidence disambiguation criterion alone would be too strict a requirement since, for nearly vertical magnetic fields, the transverse component can be less than~150~G.
We note that although some of the pixels in our data have a weaker transverse field than required by the high-confidence disambiguation criterion, this does not compromise our analysis.
Close to the disk center, radial and LOS directions are close to each other, and disambiguation does not matter as much as further away from the disk center.
Also, we used random disambiguation for weak-confidence pixels, which prevents any systematic error from arising from erroneous disambiguation.
In Appendix \ref{sec:appendix} we confirm our results using $\gamma_\mathrm{LOS}$ instead of $\gamma$, showing that the effect of disambiguation for our analysis is minuscule.

\subsection{Center-of-disk images}
\begin{figure} 
\resizebox{\hsize}{!}{\includegraphics{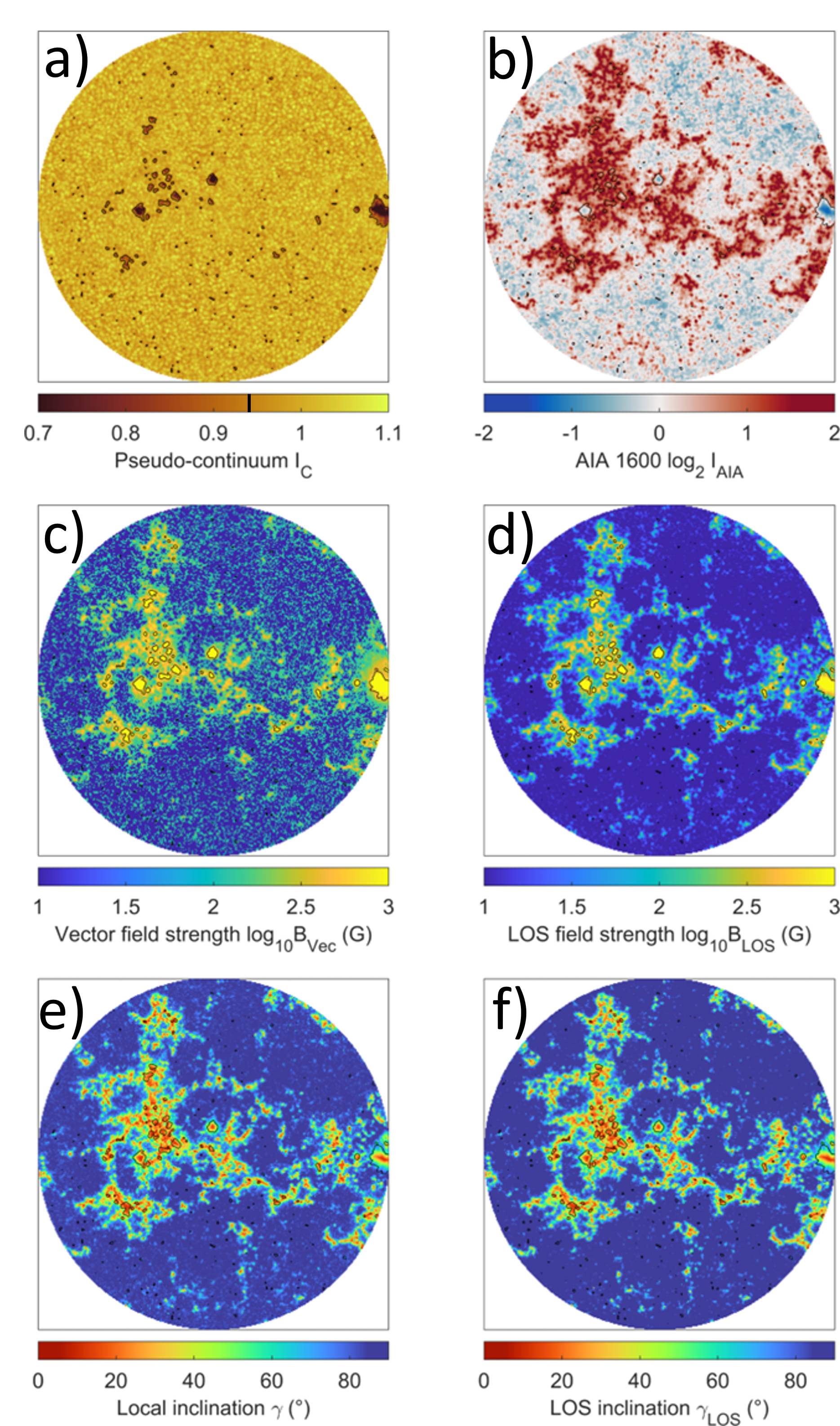}}
\caption{Center of the disk ($<0.1R_\odot{}$) on 24 August 2014. (a) HMI pseudo-continuum. (b) AIA~1600 intensity on a $\mathrm{log_2}$ scale. (c) Vector magnetic field strength on a $\mathrm{log_{10}}$ scale. (d) LOS magnetic field strength on a $\mathrm{log_{10}}$ scale. (e) Local inclination. (f) LOS inclination of the magnetic field. Black lines show the contours for the pseudo-continuum brightness level $I_\mathrm{C}~=~0.94$, which we use to define sunspot boundaries. The brightness threshold is also marked in the colorbar of panel a.}\label{fig:closeups}
\end{figure}

Figure~\ref{fig:closeups} shows the region within 0.1 $R_\odot{}$ of the solar disk center on 24 August 2014 with the contours of the $I_C < 0.94$ regions (sunspots).
There are many really small contoured regions that do not seem to correspond to sunspots or pores.
This is due to the high sunspot threshold value (0.94), which also excludes some of the darker intergranular regions.
A few missing intergranular regions do not affect our analysis but they show that we can be confident that the sunspots are excluded from the data.

\begin{figure} 
\resizebox{\hsize}{!}{\includegraphics{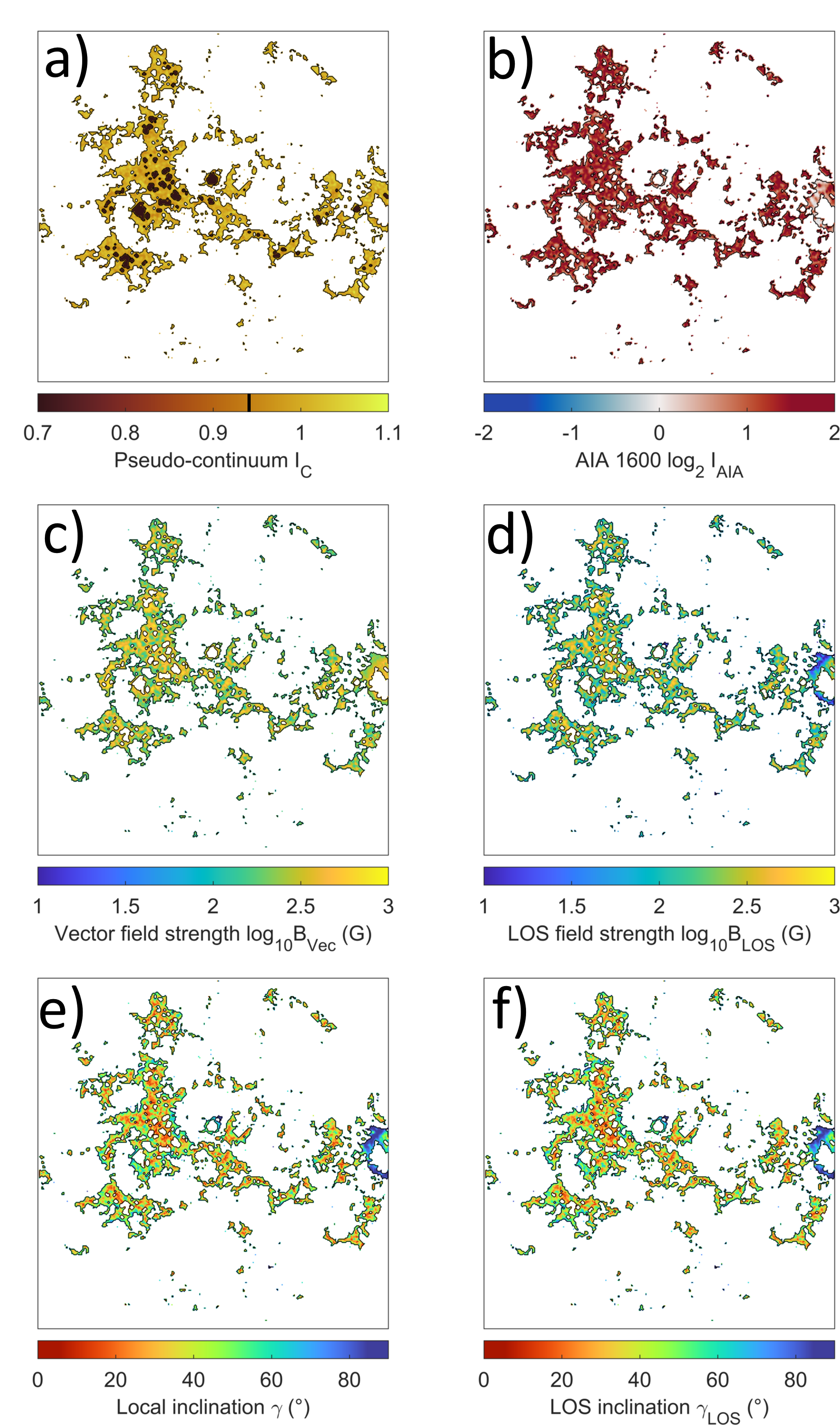}}
\caption{Center of the disk ($<0.1R_\odot{}$) on 24 August 2014 without the quiet Sun or sunspots. (a) HMI pseudo-continuum. (b) AIA~1600 intensity on a $\mathrm{log_2}$ scale. (c) Vector magnetic field strength on a $\mathrm{log_{10}}$ scale. (d) LOS magnetic field strength on a $\mathrm{log_{10}}$ scale.  (e) Local inclination. (f) LOS inclination of the magnetic field.
The sunspots are shown (in black) for the pseudo-continuum in panel a but removed from the other panels.
}\label{fig:closeupsAR}
\end{figure}

Figure~\ref{fig:closeupsAR} shows the same plots as Fig.~\ref{fig:closeups} after removing the quiet Sun and sunspots.
For reference, we show the sunspots painted black in the pseudo-continuum image of Fig.~\ref{fig:closeupsAR}a.
Figures~\ref{fig:closeups}b and ~\ref{fig:closeupsAR}b show the AIA~1600 intensity with respect to the average quiet Sun.
From Fig.~\ref{fig:closeupsAR}b it is evident that the AIA 1600 emission is strongly enhanced in active regions outside the quiet Sun and sunspots.
Figure~\ref{fig:closeups}b shows that the AIA 1600 emission is most suppressed within the large sunspots, especially the one at the right limb.

Figures~\ref{fig:closeups}c and~\ref{fig:closeups}d show the vector and LOS magnetic field strengths.
The most obvious difference between Figs.~\ref{fig:closeups}c and~\ref{fig:closeups}d is the higher noise level in the vector field measurement (Fig.~\ref{fig:closeups}c).
The differences between the vector field strength and LOS field strength in the active regions (Figs.~\ref{fig:closeupsAR}c and~\ref{fig:closeupsAR}d) are subtler.
They differ most around the large sunspot on the right limb of the image.
In this region, $B_\mathrm{LOS}$ is weaker than anywhere else in Fig.~\ref{fig:closeupsAR}d.
On the other hand, the value of $B_\mathrm{Vec}$ in this region does not differ from the rest of Fig.~\ref{fig:closeupsAR}c.
Since $B_\mathrm{LOS} \approx B_\mathrm{r}$, this difference indicates that the magnetic field in the region around the large sunspot is close to horizontal.

Figures~\ref{fig:closeups}e and~\ref{fig:closeups}f show the local and LOS inclinations for all pixels and Figs.~\ref{fig:closeupsAR}e and~\ref{fig:closeupsAR}f for active regions.
Both sets of images are almost identical due to only a small difference between $\gamma$ and $\gamma_\mathrm{LOS}$ close to the disk center.
One difference is that $\gamma$ is somewhat noisier in Fig.~\ref{fig:closeups}e than $\gamma_\mathrm{LOS}$ in Fig.~\ref{fig:closeups}f, which is probably related to the random method used in azimuthal disambiguation.
The region around the large sunspot at the right limb also shows up in Figs.~\ref{fig:closeupsAR}e and~\ref{fig:closeupsAR}f.
The magnetic field in this region is more horizontal than anywhere else  (it is almost entirely horizontal).
This is also consistent with the abovementioned difference between $B_\mathrm{Vec}$ and $B_\mathrm{LOS}$ (see Figs.~\ref{fig:closeups}c and~\ref{fig:closeups}d).

\section{AIA~1600 intensity versus magnetic field strength and inclination}\label{sec:Inclination}
\begin{figure}[!htbp]
\resizebox{\hsize}{!}{\includegraphics{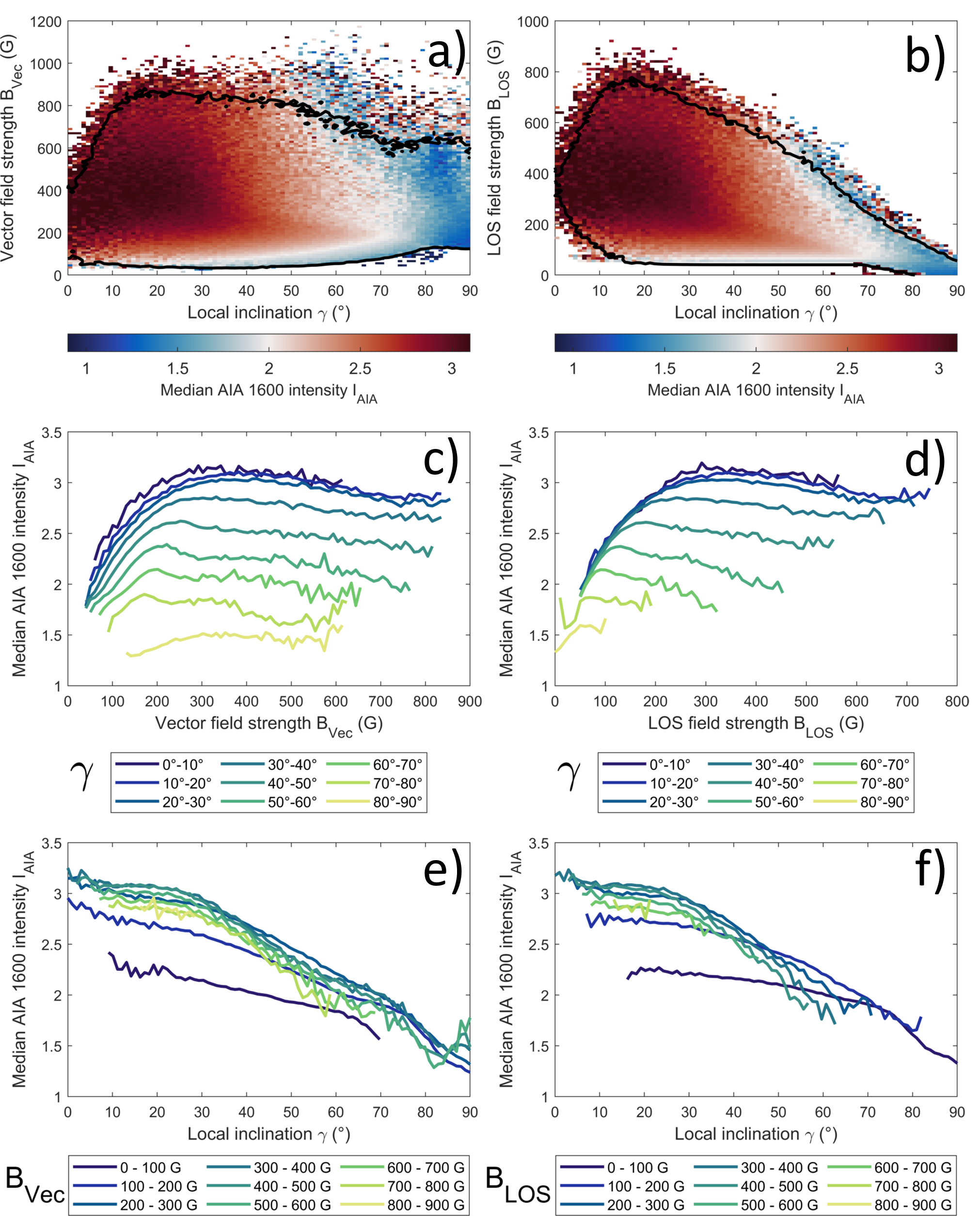}}
\caption{Median AIA~1600 intensity as a function of local inclination and magnetic field strength. (a) Median AIA~1600 intensity as a function of local inclination and $B_\mathrm{Vec}$.
The bin size is $\mathrm{\Delta} \gamma$~=~1\degree{} for local inclination and $\mathrm{\Delta} B_\mathrm{Vec}$~=~10~G for vector field strength.
The median AIA~1600 intensities are color-coded according to the colorbar.
White corresponds to the intensity level $I_\mathrm{AIA}~=~2$, i.e., twice the average level of the quiet Sun.
The black line shows the contour for bins with at least ten observations.
(b) Same as panel a, but for $B_\mathrm{LOS}$. 
(c) Median AIA~1600 intensity as a function of $B_\mathrm{Vec}$ for different inclination bins in steps of $10\degree{}$.
(d) Same as panel c, but for $B_\mathrm{LOS}$.
(e) Median AIA~1600 intensity as a function of local inclination for different fixed magnetic field strengths in steps of 100~G.
(f) Same as panel e, but for $B_\mathrm{LOS}$.
}\label{fig:AIA_vs_BandLOS_and_Inc}
\end{figure}

Figure~\ref{fig:AIA_vs_BandLOS_and_Inc}a shows the median AIA~1600 intensity as a function of local inclination $\gamma$ and vector field strength $B_\mathrm{Vec}$. This panel shows that the AIA~1600 intensity depends both on $\gamma$ and $B_\mathrm{Vec}$.
The most evident observation is that for fixed $B_\mathrm{Vec}$, AIA~1600 intensity decreases with increasing inclination for all magnetic field strengths.
Figure~\ref{fig:AIA_vs_BandLOS_and_Inc}b is the same as \ref{fig:AIA_vs_BandLOS_and_Inc}a except $B_\mathrm{LOS}$ is used instead of $B_\mathrm{Vec}$.
It is quite similar to Fig.~\ref{fig:AIA_vs_BandLOS_and_Inc}a.
However, the maximum range of $B_\mathrm{LOS}$ decreases with increasing inclination because close to the disk center $B_\mathrm{LOS} \approx B_\mathrm{r}~=~B_\mathrm{Vec}\cos{(\gamma)}$.
In Fig.~\ref{fig:AIA_vs_BandLOS_and_Inc}b, there is also a straight horizontal line from  0 to about 70\degree{}, which corresponds to the excluded quiet-Sun pixels below $B_\mathrm{LOS}~=~50$~G.
Above 60\degree, there are points below 50~G, which correspond to high-confidence disambiguation pixels.

Figure~\ref{fig:AIA_vs_BandLOS_and_Inc}c shows the median AIA~1600 intensity as a function of $B_\mathrm{Vec}$ for nine 10\degree{}-wide bins of local inclination.
These bins correspond to vertical slices through Fig.~\ref{fig:AIA_vs_BandLOS_and_Inc}a with a fixed local inclination in 10\degree{} bins in steps of 10\degree.
Median AIA~1600 intensity behaves quite similarly as a function of vector field strength for most inclinations.
Intensity increases rapidly with $B_\mathrm{Vec}$ to some peak intensity and then slowly decreases for higher field strengths.
Both the peak intensity and the field strength at which the maximum is reached decrease with the median inclination.
For fixed inclination, AIA~1600 intensity typically increases with increasing magnetic field strength until about 300-400~G for low inclination and about 150-200~G for larger inclinations.
Figure~\ref{fig:AIA_vs_BandLOS_and_Inc}d is the same as Fig.~\ref{fig:AIA_vs_BandLOS_and_Inc}c but for $B_\mathrm{LOS}$, and corresponds to vertical slices through Fig.~\ref{fig:AIA_vs_BandLOS_and_Inc}b.
In the case of $B_\mathrm{LOS}$, all median AIA~1600 intensity curves corresponding to different field inclinations follow the same common curve at low $B_\mathrm{LOS}$ strength before saturating at about 50-350~G depending on inclination.
This also leads to the fact that the peak intensities are obtained for somewhat lower field strengths in LOS measurement than in vector measurement, especially for large inclinations.

Figure~\ref{fig:AIA_vs_BandLOS_and_Inc}e shows the median AIA~1600 intensity as a function of local inclination for nine 100~G-wide bins of $B_\mathrm{Vec}$ in steps of 100~G.
These bins correspond to horizontal slices through Fig.~\ref{fig:AIA_vs_BandLOS_and_Inc}a for a fixed $B_\mathrm{Vec}$ value.
Figure~\ref{fig:AIA_vs_BandLOS_and_Inc}e shows that the median AIA~1600 intensity decreases with increasing local inclination for all magnetic field strengths.
The highest median AIA~1600 intensity is associated with close-to-vertical fields, but interestingly, the median intensity increases with increasing magnetic field strength only for low field strengths  ($B_\mathrm{Vec} < 300~\mathrm{G}$) and is fairly constant for higher field strengths.
For small inclinations of about $\gamma < 45\degree{}$, the median intensity for $B_\mathrm{Vec} > 200~\mathrm{G}$ remains at a fairly similar level and decreases with inclination quite similarly.
Above 45\degree{}, the median intensity decreases slightly faster with increasing inclination for high magnetic field strengths, and at around 75\degree{} the slope steepens further.
Figure~\ref{fig:AIA_vs_BandLOS_and_Inc}f is the same as Fig.~\ref{fig:AIA_vs_BandLOS_and_Inc}e but for $B_\mathrm{LOS}$, and corresponds to horizontal slices through Fig.~\ref{fig:AIA_vs_BandLOS_and_Inc}b.

\begin{figure} 
\resizebox{\hsize}{!}{\includegraphics{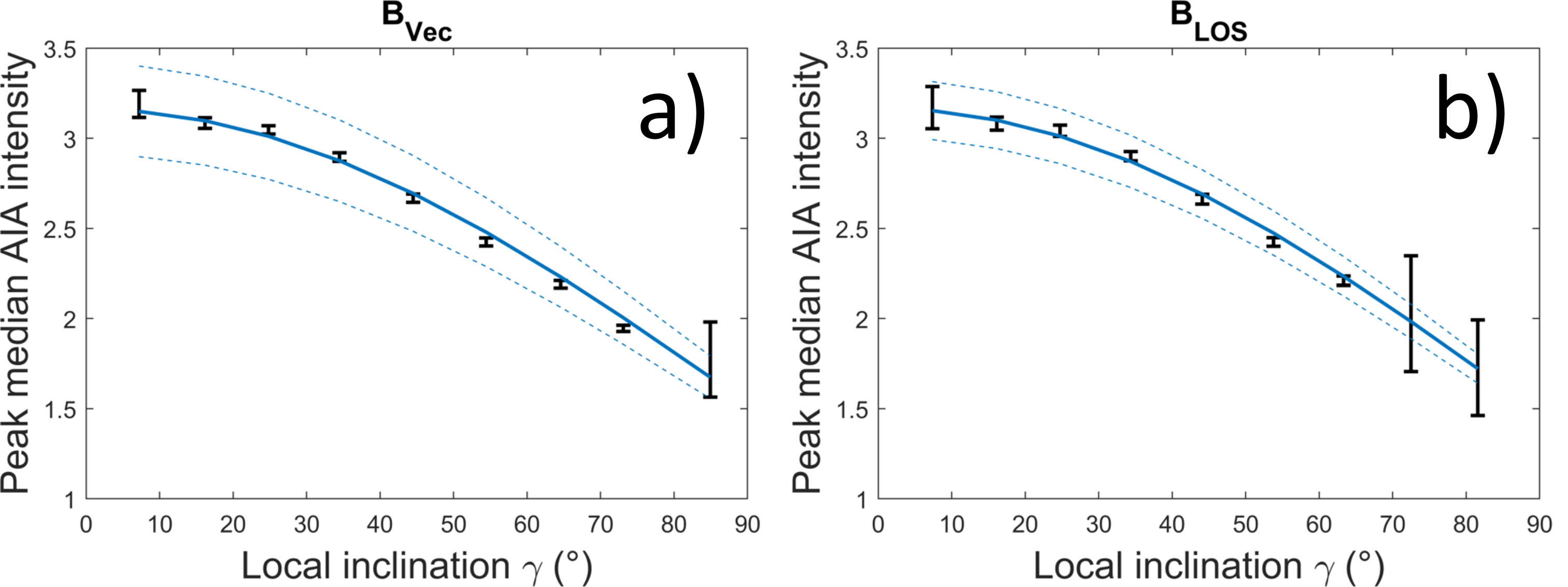}}
\caption{Peak median AIA 1600 intensities from Figs.~\ref{fig:AIA_vs_BandLOS_and_Inc}c and \ref{fig:AIA_vs_BandLOS_and_Inc}d as a function of local inclination. Error bars correspond to twice the bootstrap standard error. The solid lines show the best cosine fits to peak intensities. Dashed lines show 95\% confidence bounds of the cosine fit.
}\label{fig:peakemission}
\end{figure}

Figure \ref{fig:peakemission} shows the peak intensities of the median intensity curves depicted in Figs.~\ref{fig:AIA_vs_BandLOS_and_Inc}c and \ref{fig:AIA_vs_BandLOS_and_Inc}d as a function of inclination.
The error bars show twice the bootstrap standard error for the peak median intensity.
We calculated the bootstrap error by resampling the AIA intensities within each peak median intensity bin 10~000 times.
The bootstrap standard error is quite small for all but the most horizontal fields due to the large number of data points.
The lines show the best-fit cosine function with 95\% confidence bounds.
The best-fit parameters for both $B_\mathrm{Vec}$ and $B_\mathrm{LOS}$ binning are shown in Table \ref{table:peakemission}.
Figure \ref{fig:peakemission} together with Table \ref{table:peakemission} demonstrate that the peak median AIA intensity closely follows the cosine of local inclination.
We also note that the two fits corresponding to different magnetic field binnings agree well with each other.
Both fits show that peak intensity is $I_\mathrm{Peak} = 3.17$ at a perfectly radial field and $I_\mathrm{Peak} = 1.40$ (1.36) at a perfectly horizontal field.

\begin{table}
\caption{Regression coefficients of the model $I_\mathrm{Peak} = \beta_0 + \beta_1\cos{\gamma}$.}              
\label{table:1}      
\centering                                      
\begin{tabular}{c | c  c | c}          
\hline\hline                        
  &  $\beta_0$ & $\beta_1$ & $R^2$ \\    
\hline                                   
    $B_\mathrm{Vec} $ & $1.40 \pm 0.05$ & $1.77 \pm 0.08$  & 0.998 \\      
    $B_\mathrm{LOS} $ & $1.36 \pm 0.08$ & $1.81 \pm 0.11$  & 0.996 \\
\hline                                             
\end{tabular}\label{table:peakemission}
\end{table}

\section{Boundary layers and distributions}\label{sec:boundary}
Figure~\ref{fig:closeupsAR} shows that the AIA 1600 intensity and magnetic field strength and inclination are not uniformly distributed within the active regions.
The magnetic field strength tends to increase from the boundary toward the inner regions, while the inclination decreases.
In this section we study these changes and their consequences in more detail.

\subsection{Median masks}
\begin{figure} 
\resizebox{\hsize}{!}{\includegraphics{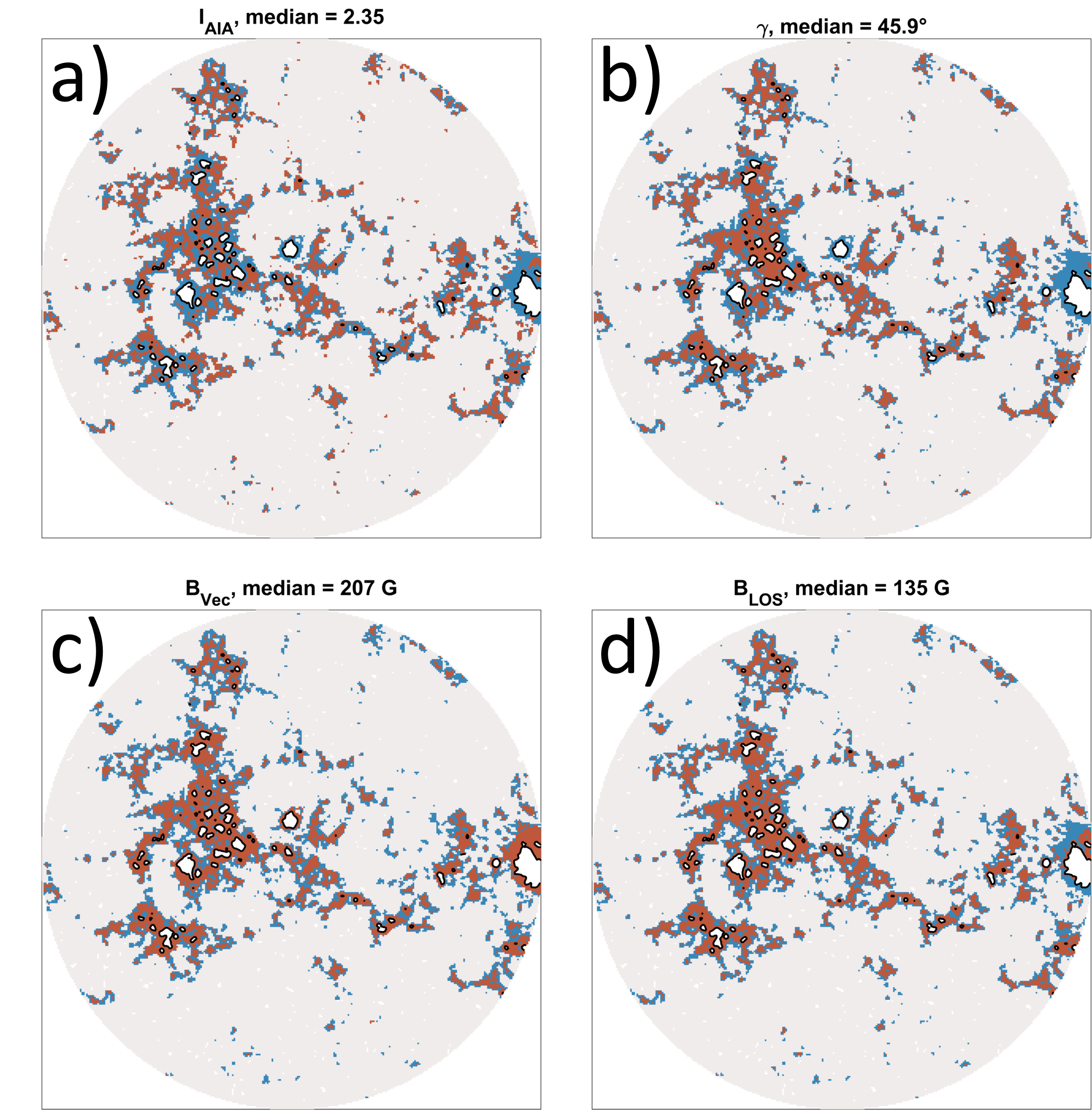}}
\caption{Same as panels b-e of Fig.~\ref{fig:closeupsAR}, but with pixels above and below the median differentiated by color.
(a) Pixels above (red) and below (blue) the median AIA~1600 intensity $I_\mathrm{AIA}~=~2.35$. (b) Pixels below (red) and above (blue) the median inclination $\gamma~=~45.9\degree{}$. 
(c) Pixels above (red) and below (blue) the median $B_\mathrm{Vec}~=~207$~G. (d) Pixels above (red) and below (blue) the median $B_\mathrm{LOS}~=~135$~G.
}\label{fig:closeupsmasks}
\end{figure}

This difference between boundary and core regions is made clearer in Fig.~\ref{fig:closeupsmasks}, which shows the same region as Fig.~\ref{fig:closeupsAR} but with different colors marking the pixels below and above the median.
Figure~\ref{fig:closeupsmasks} reveals that all four quantities are organized similarly, with less bright, weaker, and high-inclination pixels appearing dominantly at the boundaries of active regions, and the high-intensity, stronger, more vertical pixels in the core.

The region around the large sunspot on the right limb solar disk center stands out in Fig.~\ref{fig:closeupsmasks}.
Lower-intensity high-inclination pixels dominate the region around the sunspot.
This region also shows the largest difference between $B_\mathrm{Vec}$ and $B_\mathrm{LOS},$ as discussed above.
While the $B_\mathrm{LOS}$ in this region is almost entirely below the median value, the $B_\mathrm{Vec}$ is almost entirely above its median.
This tells us that the region surrounding the sunspot has a strong but highly horizontal magnetic field with a below-median AIA~1600 intensity.
This region may correspond to the so-called superpenumbra, the chromospheric extension of sunspot penumbrae that typically hosts highly inclined magnetic fields \citep{Zhang1996,Solanki2003,Sobotka2013}.
As can be seen from Fig.~\ref{fig:closeups}a, this region does not manifest as suppressed pseudo-continuum radiation.

\subsection{Boundary layers}
Here we study how the distributions of AIA~1600 intensity, vector field strength, and local inclination behave spatially.
For this purpose, we define a special type of active region, an activity cluster, as a set of pixels that are attached by either their sides or corners.
We use the term "activity cluster" instead of "active region" (or "plage") since an active region is usually associated with a large-scale structure observed in lower-resolution measurements, whereas higher-resolution measurements consist of many smaller objects.
We removed all activity clusters with fewer than 1000 pixels (area of 360 $\mathrm{arcsec^2}$) and activity clusters touching the limb of the studied $0.1R_\odot{}$ region.
We studied how the AIA 1600 Å intensity, magnetic field strength, and magnetic field inclination of activity clusters behave as a function of distance from the activity cluster perimeter.
For our analysis, we defined the concept of the N-boundary, where N is an integer and refers to the Nth layer of an activity cluster starting from the border of an activity cluster.
The process of defining the N-boundaries resembles the peeling of an onion.
The 1-boundary is the outmost one-pixel-wide boundary of an activity cluster (i.e., its perimeter).
The 2-boundary is the second outmost one-pixel-wide boundary of an activity cluster (i.e., the new perimeter after peeling away the 1-boundary).
Likewise, the N-boundary is the Nth such layer of an activity cluster after peeling off N-1 layers.
For simplicity, we discuss here only the outer boundary of the activity clusters, not the inside boundary of the holes inside activity clusters which we fill before the boundary calculation.
The holes inside activity clusters are formed mainly when excluding the sunspots, while filling holes is done only to determine the (outer) boundary locations.
Including sunspots when calculating these boundaries is important because they are naturally part of the larger active region that produces them.

\begin{figure}[!htbp]
\resizebox{\hsize}{!}{\includegraphics{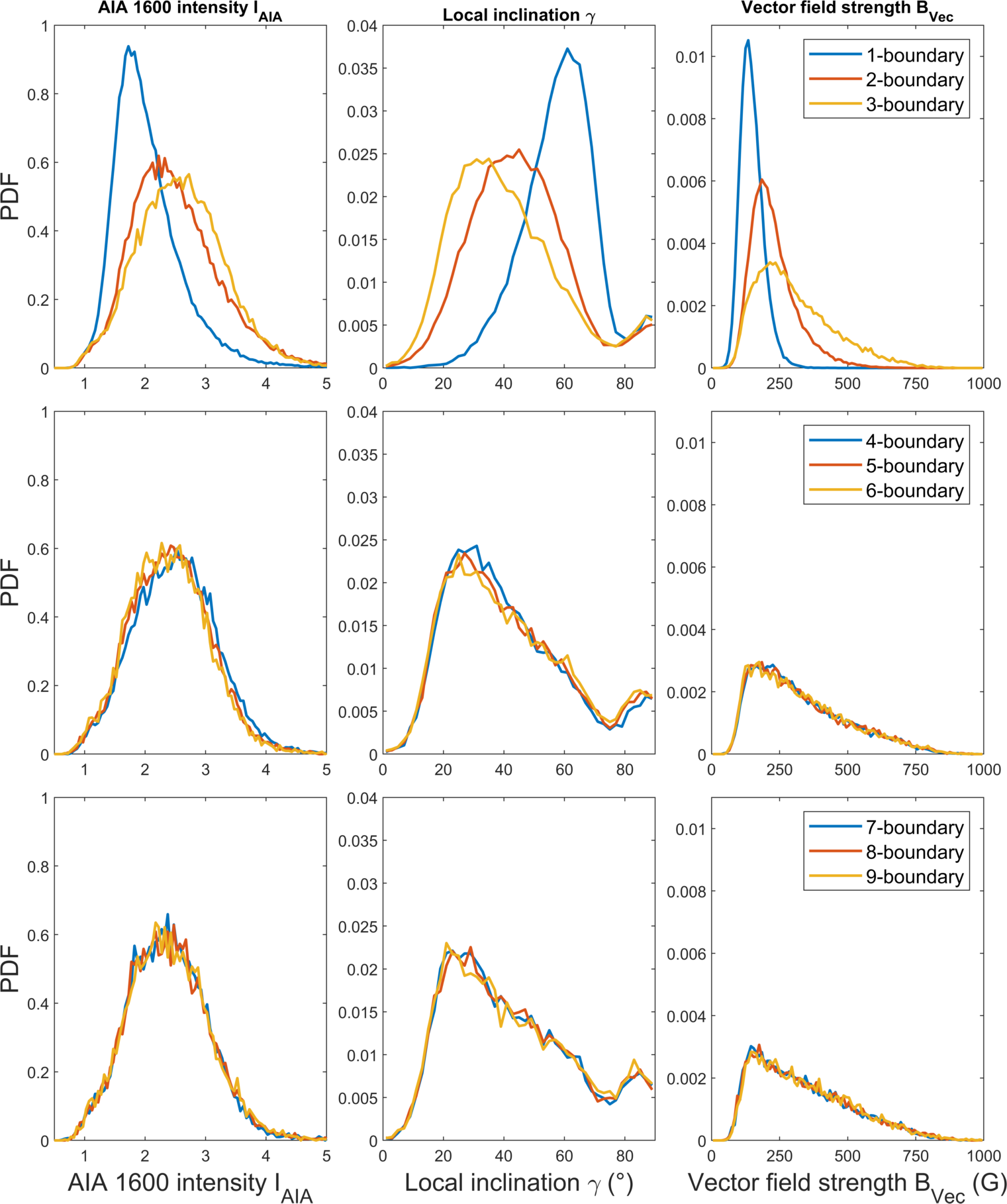}}
\caption{N-boundary distributions for A $>$ 1000 px activity clusters. Left column: AIA~1600 intensity. Middle column: Local inclination, $\gamma$. Right column: Vector field strength, $B_\mathrm{Vec}$. The first row shows the distributions for the 1-boundary (blue), 2-boundary (orange), and 3-boundary (yellow), the second row for the 4-boundary, 5-boundary, and 6-boundary, and the third row for the 7-boundary, 8-boundary, and 9-boundary.
}\label{fig:boundarydistributions}
\end{figure}

Figure~\ref{fig:boundarydistributions} shows the probability distributions of AIA~1600 intensity $I_\mathrm{AIA}$ (left column), local inclination $\gamma$ (middle column), and vector magnetic field strength $B_\mathrm{Vec}$ (right column) distributions for the first nine boundary layers.
The first row shows the distributions for the first three boundary layers, the second for layers 4 to 6, and the third for layers 7 to 9.
The first row of Fig.~\ref{fig:boundarydistributions} shows that the distributions change quite a lot between the first three boundary layers.
The 1-boundary (original perimeter) in particular is both dimmer and more inclined and has a weaker magnetic field than all other layers.
The intensity of emission increases, and the magnetic field becomes more radial and stronger when moving from the original perimeter to the second and third boundary.
However, the differences between consecutive boundary layers soon vanish as shown by the second row of Fig.~\ref{fig:boundarydistributions}.
AIA~1600 intensity distributions are already quite stable from the 2-boundary onward; beyond the 4-boundary, the distributions are practically indistinguishable between the different layers.
The distributions of local inclination and vector field strength show large differences between the first three boundary layers.
However, even the $B_\mathrm{Vec}$ distributions become practically indistinguishable beyond the 3-boundary.
Overall, Fig.~\ref{fig:boundarydistributions} shows that the emission is brighter and the magnetic field is stronger and more radial inside the activity cluster than at the first two boundaries.

Inclination distributions in Fig.~\ref{fig:boundarydistributions} are bimodal, with the first peak around 25\degree{}, the second peak close to 90\degree,{} and a local minimum at 76\degree.
The second peak, corresponding to almost completely horizontal magnetic fields, is likely related to suggested superpenumbral regions \citep{Zhang1996,Solanki2003,Sobotka2013} appearing around the largest sunspots, as discussed above (see also Fig.~\ref{fig:closeupsmasks}).
We find that 85.3\% of the $\gamma > 76\degree$ pixels are connected to $I_\mathrm{C} < 0.94$ sunspot pixels either directly or as neighbors connected by other $\gamma > 76\degree$ pixels.
Also, when experimenting with extending the sunspot regions, we found that the excluded pixels were primary in the $\gamma > 76\degree$ range.

\begin{figure}[!htbp]
\resizebox{\hsize}{!}{\includegraphics{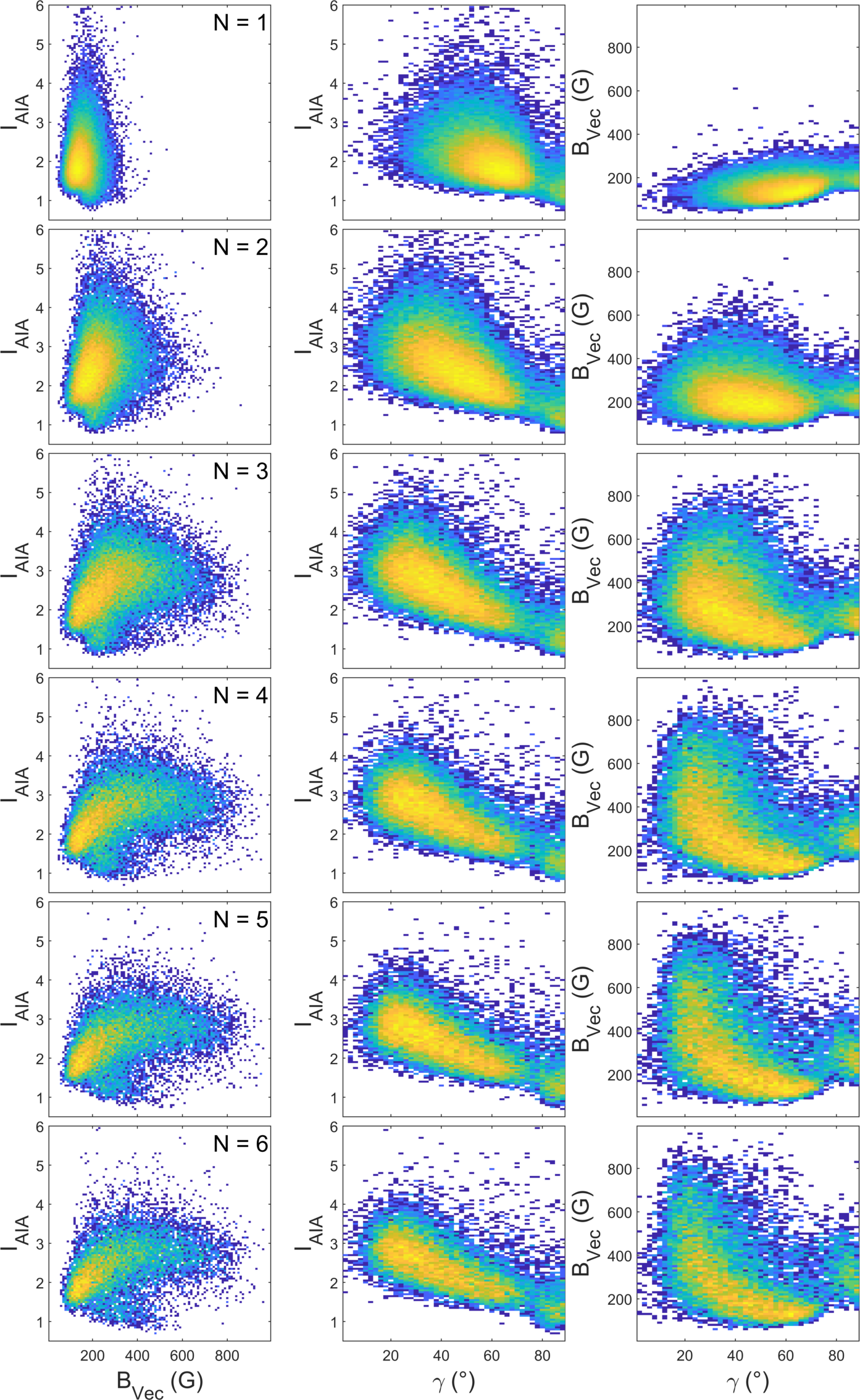}}
\caption{Bivariate boundary distributions for A $>$ 1000 px activity clusters. Left column: $B_\mathrm{Vec}-I_\mathrm{AIA}$ distributions.\ Middle column: $\gamma-I_\mathrm{AIA}$ distributions. Right column: $\gamma-B_\mathrm{Vec}$ distributions. The first row shows the distributions for the 1-boundary, the second for the 2-boundary, etc.
}\label{fig:boundarydistributions2D}
\end{figure}

Figure~\ref{fig:boundarydistributions2D} provides detailed information on how AIA~1600 intensity, vector field strength, and local inclination change from the original perimeter toward the inner boundaries by showing the mutual relations between the three bivariate histograms for the three parameters for six outermost boundary layers.
The columns of Fig.~\ref{fig:boundarydistributions2D} show the histograms for $B_\mathrm{Vec}-I_\mathrm{AIA}$ (left column), $\gamma-I_\mathrm{AIA}$ (middle column), and $\gamma-B_\mathrm{Vec}$ (right column) while rows~1~to~6 correspond to boundary layers~1~to~6.
Again, the distributions show noticeable differences between the first few boundary layers.
The distributions involving the vector field strength $B_\mathrm{Vec}$ (left and right column) change noticeably between the first three boundary layers, but the $\gamma-I_\mathrm{AIA}$ distribution (middle column) only between the first and second.
Indeed, the shape of $\gamma-I_\mathrm{AIA}$ distribution is similar for all boundary layers except for the 1-boundary.
The bimodal inclination distribution seen in Fig.~\ref{fig:boundarydistributions} can also be seen in the two bivariate distributions of Fig.~\ref{fig:boundarydistributions2D} involving the local inclination (middle and right panels).
Interestingly, in the middle column of Fig.~\ref{fig:boundarydistributions2D}, AIA~1600 intensity values seem to change continuously from vertical down to the most horizontal superpenumbral distribution.
As a comparison, in the right column of Fig.~\ref{fig:boundarydistributions2D} the most horizontal population appears as clearly distinct from the rest of the pixels for $N > 3$ boundaries.
This horizontal population is also visible in the left column of Fig.~\ref{fig:boundarydistributions2D} between 200~G and 400~G and below I~=~2 as an extension of the high-density region.

\begin{figure}[!htbp]
\resizebox{\hsize}{!}{\includegraphics{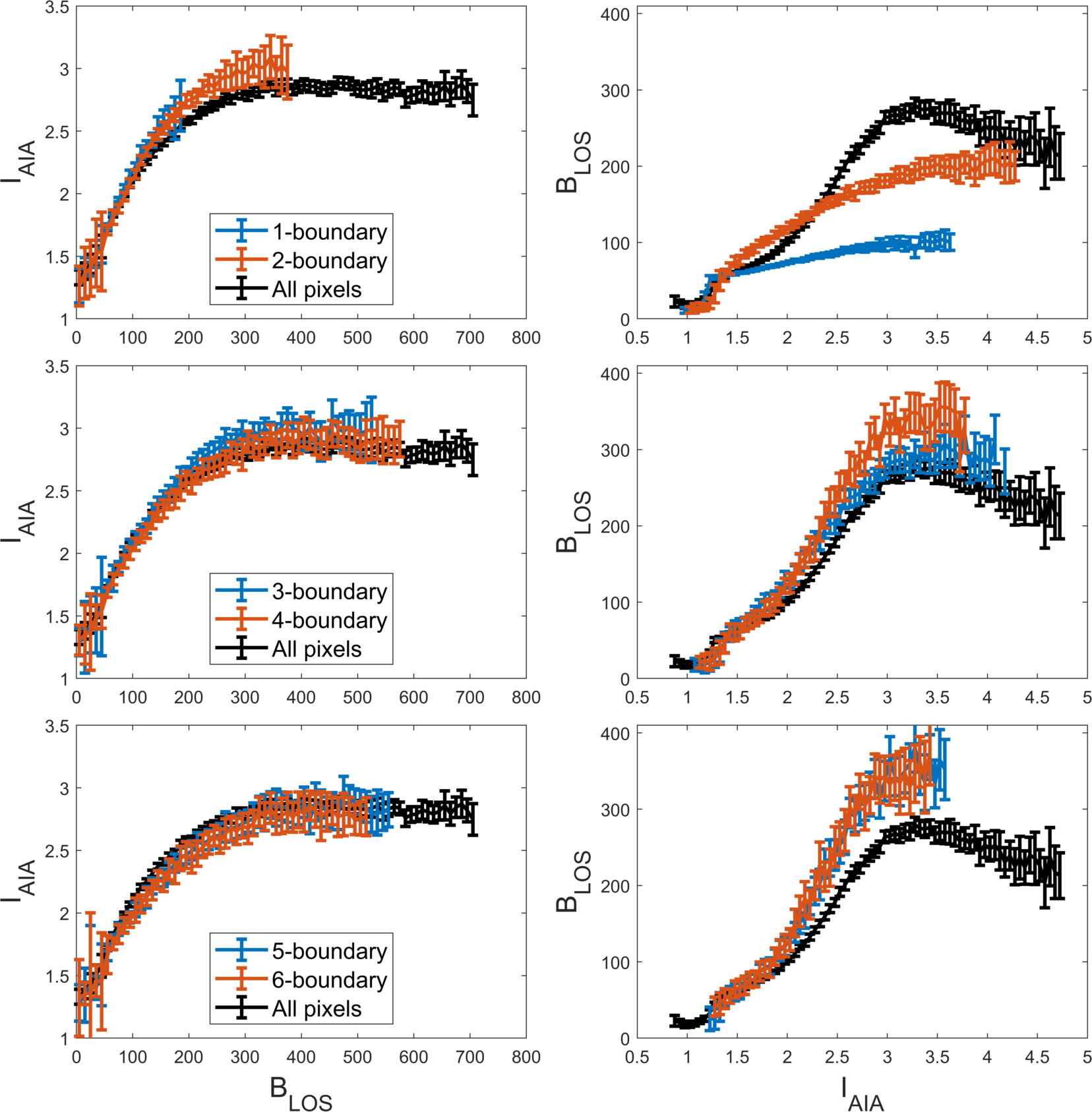}}
\caption{Boundary relations for A $>$ 1000 px activity clusters. Left column: Median AIA~1600 intensity as a function of $B_\mathrm{LOS}$ binned in steps of $\mathrm{\Delta} B_\mathrm{Vec}$ = 10~G. Right column:  Median $B_\mathrm{LOS}$ intensity as a function of AIA~1600 binned in steps of $\mathrm{\Delta} I_\mathrm{AIA}$ = 0.05. The first row shows the median relations for the 1-boundary (blue) and the 2-boundary (orange), the second for the 3-boundary and the 4-boundary, and the third for the 5-boundary and the 6-boundary. Black lines show the median relations calculated from all activity cluster pixels.
Error bars show twice the bootstrap standard error.
The bins that have fewer than 100 data points are omitted.
}\label{fig:boundaryrelations}
\end{figure}

The fact that the bivariate histograms of, for example, AIA~1600 intensity and vector field strength differ from each other for the first few boundary layers has implications for the relation between the two quantities. 
Figure~\ref{fig:boundaryrelations} (left column) shows the median AIA~1600 intensity as a function of binned $B_\mathrm{LOS}$ and Fig.~\ref{fig:boundaryrelations} (right column) the median $B_\mathrm{LOS}$ as a function of binned AIA~1600 intensity for boundary layers 1--6.
Here we study $B_\mathrm{LOS}$ instead of $B_\mathrm{Vec}$ because the early systematic magnetic field observations that are important for reconstructing the magnetic field over the past century \citep{Virtanen2019} measured LOS magnetic field strength \citep{Pevtsov2021Longterm}.
Also, most of the earlier studies relating chromospheric emissions to photospheric magnetic fields \citep[][]{Skumanich1975,Schrijver_1989,HarveyWhite1999,Ortiz_2005,Rezaei_2007,Loukitcheva2009,Pevtsov.etal2016,Chatzistergos2019,Tahtinen2022} typically employ $B_\mathrm{LOS}$.
The black line in Fig.~\ref{fig:boundaryrelations} shows the median relation calculated using all activity cluster pixels; it thus remains the same for all three panels in one column.
Figure~\ref{fig:boundaryrelations} shows that the relation between the AIA~1600 intensity and the LOS field strength, $B_\mathrm{LOS}$, changes between the first few boundary layers of an activity cluster.
However, there is a big difference in the ordering of binning, that is, in how much the $B_\mathrm{LOS}-I_\mathrm{AIA}$ and $I_\mathrm{AIA}-B_\mathrm{LOS}$ relations change.
While the $B_\mathrm{LOS}-I_\mathrm{AIA}$ relation (left panels of Fig.~\ref{fig:boundaryrelations}) changes only a little, the $I_\mathrm{AIA}-B_\mathrm{LOS}$ relation (right panels of Fig.~\ref{fig:boundaryrelations}) shows significant differences between different boundaries.
Near the original perimeter (the 1-boundary), the median magnetic field is much weaker for a fixed AIA~1600 intensity bin than for the inner boundaries.
Beyond the 3-boundary these distributions resemble each other closely.
The median relation of all activity cluster pixels (black curve Fig.~\ref{fig:boundaryrelations}) overestimates the magnetic field strength near the boundary but underestimates it within the activity cluster interior.
These differences between the $B_\mathrm{LOS}-I_\mathrm{AIA}$ and $I_\mathrm{AIA}-B_\mathrm{LOS}$ relations stem from the dominance of weak fields close to the activity cluster perimeter visible in the upper left panel of Fig.~\ref{fig:boundarydistributions2D}.
The large variation in AIA~1600 intensity for fixed $B_\mathrm{LOS}$ (and $B_\mathrm{Vec}$) mostly cancels out when averaged for a fixed magnetic field bin.
However, when the magnetic field is averaged for fixed AIA~1600 intensity bins the relative fraction of weak fields varies greatly between the first three boundaries.

\section{Discussion}\label{sec:discussion}
We studied the interdependence of AIA~1600 intensity, magnetic field strength ($B_\mathrm{Vec}$ and $B_\mathrm{LOS}$), and magnetic field inclination in moderate field strength regions (active regions) outside the sunspots over a period from 1~March~2014 to~9~June~2017.
These regions are closely related to chromospheric plages seen in AIA~1600 emission and many other wavelength bands and emission lines \citep[see, e.g.,][]{Schrijver_1989,Barczynski2018,Tahtinen2022}.
We excluded the sunspots from the data using a pseudo-continuum intensity lower threshold of $I_C~=~0.94$.
The exclusion of sunspots is important because the relation between emissions and magnetic field strength in sunspots is known to differ from that within moderate field active regions \citep{Solanki2003}.
We also removed the quiet-Sun pixels by requiring that the pixels to be analyzed have either a $B_\mathrm{LOS}$ that exceeds 50 G or a high-confidence vector field disambiguation, meaning that the transverse component of the field exceeds the vector noise level ($\approx 100~\mathrm{G}$) by 50~G.

\subsection{Effect of inclination}
Figure~\ref{fig:AIA_vs_BandLOS_and_Inc} shows how the median AIA~1600 intensity depends on the magnetic field strength ($B_\mathrm{Vec}$ and $B_\mathrm{LOS}$) and on local inclination.
AIA~1600 emission increases with increasing field strength until reaching a peak intensity at a cutoff field strength, above which the emission saturates and mostly decouples from the magnetic field strength.
This increase and subsequent saturation of chromospheric emissions as a function of magnetic flux density have been observed in many earlier studies \citep{Skumanich1975,Schrijver_1989,HarveyWhite1999,Ortiz_2005,Rezaei_2007,Loukitcheva2009,Pevtsov.etal2016,Kahil_2017,Kahil2019,Chatzistergos2019,Tahtinen2022}.
The intensity of chromospheric emissions, Ca II K in particular, has been related to the cross-sectional area of magnetic flux tubes at their emission height \citep{Schrijver_1989,Solanki1991,Barczynski2018,Kahil2019}.
The saturation follows when the atmosphere at the emission height is completely filled with expanding magnetic flux tubes so that their cross-sectional area within a resolution element cannot increase further even if more magnetic flux is brought into the region.

An important result of this paper is that the relation between AIA 1600 intensity and magnetic field strength depends on magnetic field inclination as shown in Figs.~\ref{fig:AIA_vs_BandLOS_and_Inc}c and~\ref{fig:AIA_vs_BandLOS_and_Inc}d.
The peak intensity at which the emission saturates decreases with increasing inclination: close-to-radial magnetic fields emit more radiation than more horizontal fields of the same strength.
In addition to peak intensity, the cutoff field strength (field strength at the peak emission) decreases from about 400~G to less than 200~G for $B_\mathrm{Vec}$ (from about 400~G to less than 200~G for $B_\mathrm{LOS}$) with increasing inclination.
Emissions from regions with more vertical magnetic fields reach their peak intensity at higher field strengths.
Above the cutoff field strength, the emission is mainly governed by inclination.
Figures~\ref{fig:AIA_vs_BandLOS_and_Inc}e and~\ref{fig:AIA_vs_BandLOS_and_Inc}f show how the median intensity decreases with increasing local inclination for different bins of magnetic field values.
For all magnetic field strengths, the highest AIA~1600 intensity is obtained for the most vertical fields.
Above about 200~G, the curves corresponding to different magnetic field strengths closely follow each other, indicating the aforementioned saturation and decoupling of emission from the field strength.

Figure ~\ref{fig:peakemission} and Table~\ref{table:peakemission} demonstrate that the peak median intensity closely depends on the cosine of local inclination.
The peak emission from purely radial fields is more than twice the emission from horizontal fields.
The constant term $\beta_0$ of the regression model (see Table \ref{table:peakemission}) represents the part of emission that is not affected by the inclination.
The value of the constant term, about 1.36--1.40, is well above the average intensity $I_0 = 1$ of the quiet Sun.
This means that the emission, at least at the peak level, consists of two components, with the $\beta_0$ term forming the base level of emissions for a sufficiently strong field, and $\beta_1$  denoting the level of emissions controlled by inclination (i.e., the direction of magnetic field).

There is a clear difference in the dependence of the median AIA~1600 intensity on $B_\mathrm{Vec}$ versus $B_\mathrm{LOS}$ in Figs.~\ref{fig:AIA_vs_BandLOS_and_Inc}c-~\ref{fig:AIA_vs_BandLOS_and_Inc}d.
For $B_\mathrm{Vec}$, the intensity curves for different inclinations are almost independent of each other.
However, for $B_\mathrm{LOS}$ the intensity curves associated with different field inclinations follow a common curve until inclination-dependent cutoff field strength.
This difference shows that the observed emission primarily depends on $B_\mathrm{LOS}$ rather than $B_\mathrm{Vec}$.
Indeed, multiplying the vector field grid values in Fig.~\ref{fig:AIA_vs_BandLOS_and_Inc}c by $\cos{\gamma}$ brings the intensity curves together similarly to Fig.~\ref{fig:AIA_vs_BandLOS_and_Inc}d.

We found a simple model that can produce the intensity curves of Figs. \ref{fig:AIA_vs_BandLOS_and_Inc}c and \ref{fig:AIA_vs_BandLOS_and_Inc}d.
Assume that there is a magnetic flux tube of total flux $\phi$ located at the center of the image pixel with the field strength following the Gaussian function 
\begin{equation}\label{eq1}
B(r) = B_0 \mathrm{e}^{-\frac{r^2}{r_0^2}},
\end{equation}
where $B_\mathrm{0}$ is the field strength at the center of the flux tube and $r$ is the distance from the flux tube center.
The decay length, $\mathrm{r_0}$, was obtained in terms of $\mathrm{B_0}$ and $\phi$ by integrating Eq. 1 from zero to infinity:
\begin{equation}\label{eq:TubeRadius}
r_0 = \sqrt{\frac{\phi}{\mathrm{\pi} B_0}}.
\end{equation}
Since it is hinted by Fig.~\ref{fig:AIA_vs_BandLOS_and_Inc} that it is the LOS field strength rather than vector field strength that governs AIA~1600 emission, we assumed that the emission is proportional to some power $\alpha$ of the LOS field strength $B_\mathrm{LOS} = B \cos{\gamma}$.
The exponent $\alpha$ was introduced because chromospheric emissions are found to follow a power law of magnetic field strength.
The modeled AIA~1600 intensity was then obtained by averaging over the image pixel:
\begin{equation}
I_\mathrm{AIA} \propto \mathrm{\frac{1}{A_{px}}} \int_A (B\cos{\gamma})^\mathrm{\alpha} \,dA = \frac{(B_0 \cos{\gamma})^\mathrm{\alpha}}{\mathrm{A_{px}}} \int_A \mathrm{e}^{-\alpha\frac{r^2}{r_0^2}} \,dA. 
\end{equation}

\begin{figure}
\resizebox{\hsize}{!}{\includegraphics{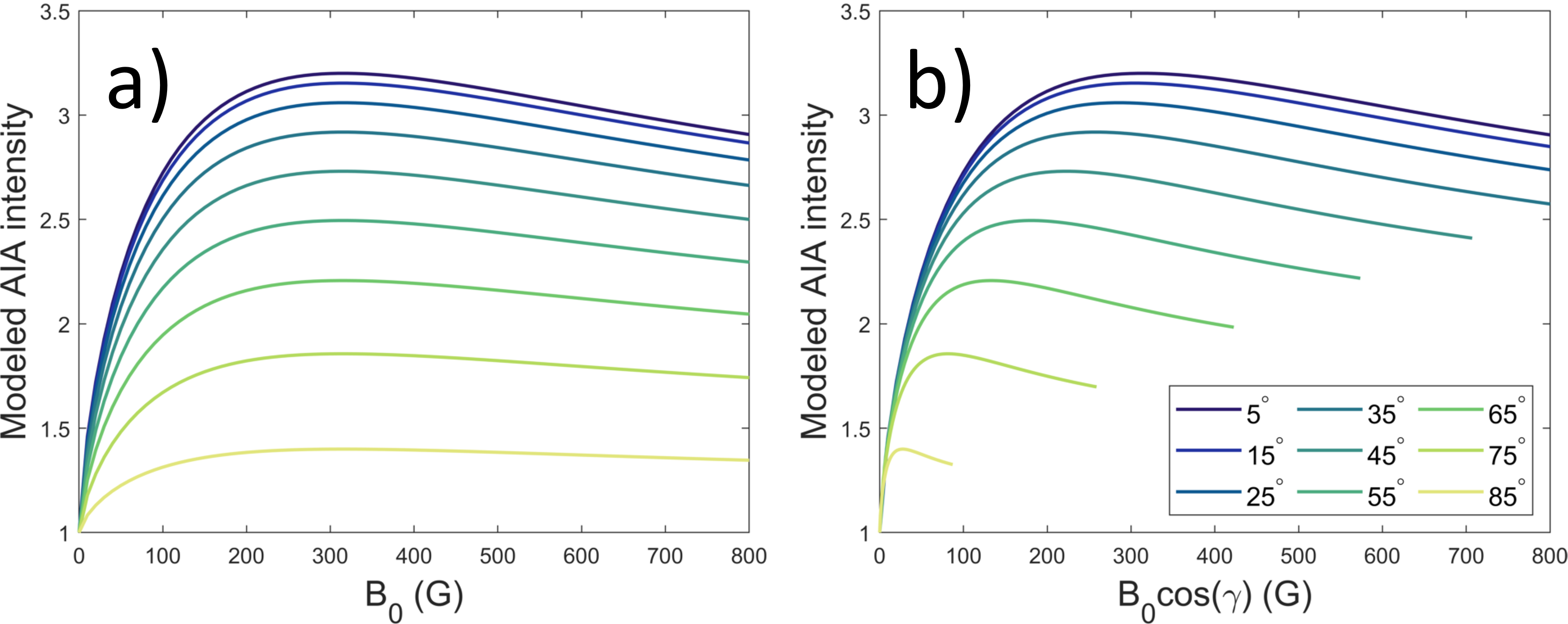}}
\caption{Modeled AIA~1600 intensity as a function of $B_0$ (panel a) and $B_0\cos{\gamma}$ (panel b) for different inclinations.}\label{fig:ModelledEmission}
\end{figure}

Figure~\ref{fig:ModelledEmission} shows the modeled intensities as a function of $B_0$ (panel a) and $B_0\cos{\gamma}$ (panel b) for different inclinations with $B_0$ ranging from 0 to 1000~G and with $\alpha~=~0.7$ and $\phi~=~2~\times~10^{17}$~Mx, which correspond to typical values of the power-law exponent and photospheric flux tube.
We scaled the modeled emission to the same level with $I_\mathrm{AIA}$.
Interpreting $B_0$ as the vector field strength and $B_0\cos{\gamma}$ as the LOS field strength measured by the magnetograph, the model produces intensity curves that closely resemble those of Fig.~\ref{fig:AIA_vs_BandLOS_and_Inc}.
Like for $B_\mathrm{Vec}$ in Fig.~\ref{fig:AIA_vs_BandLOS_and_Inc}c, intensity curves for $B_0$ are almost independent of each other, while for $B_0\cos{\gamma}$ they follow a common curve until some inclination-depended cutoff strength like for $B_\mathrm{LOS}$ in Fig.~\ref{fig:AIA_vs_BandLOS_and_Inc}d.
Interpreting $B_0$ as a measured vector field strength could be justified if the magnetic field measurement is more sensitive to strong fields within the pixel so that the measured value is more representative of maximum field strength instead of average. 

The relation between AIA~1600 intensity and inclination may also be related to the results of \citet{Rajaguru2019}, who study the propagation of magnetoacoustic gravity waves (see \citealp{Yelles2023} and references therein) from the photosphere to the emission heights of AIA 1700 (340 km) and AIA~1600 (430 km).
They find an increase in the phase travel times of waves propagating along more inclined magnetic fields, indicating propagation beyond the emission height of AIA~1600 due to reduced cutoff frequency $\nu_{\mathrm{co}} \propto \cos{\gamma}$.
\citet{Rajaguru2019} interpret their results to indicate that while the low-frequency waves propagating along more horizontal lines could propagate higher to the atmosphere due to reduced cutoff frequency, the low-frequency waves propagating along more vertical lines evolve to a shock at lower chromospheric heights.
Curiously, similarly to this cutoff frequency, we find that the AIA 1600 peak emission is proportional to $\cos{\gamma}$.
The decreasing AIA~1600 intensity with increasing inclination could be related to the result of \citet{Rajaguru2019} where the shock-formation height increases with inclination, heating plasma in a rarer environment, thus emitting less.
Our observation that the highly inclined fields with reduced emission are typically found on the boundaries of activity clusters also agrees with previous studies of wave propagation in the solar atmosphere.
\citet{Jefferies2006} showed that the boundaries of supergranular cells acted as "portals" through which low-frequency magnetoacoustic waves could propagate into the chromosphere.
Similarly, \citet{deWijn2009} recognized that low-frequency waves could leak into the chromosphere only along the inclined magnetic fields found on the periphery of plage.

\subsection{Boundary regions}
We also analyzed how the distributions of AIA 1600 intensity, magnetic field strength, and local inclination change when moving from the outer boundary of activity clusters toward their center.
Activity cluster is defined here as a set of adjacent pixels connected by either their sides or corners.
Figure~\ref{fig:boundarydistributions} shows that there are notable differences between the original perimeter (the 1-boundary) and inner layers (N-boundaries with $N\geq2$) of activity clusters.
The magnetic field tends to be weaker and more horizontal and the AIA~1600 intensity is weaker at the perimeter of an activity cluster than in deeper layers.
These differences are made clearer in Fig.~\ref{fig:boundarydistributions2D}, which shows how the bivariate distributions of AIA~1600 intensity, magnetic field strength, and local inclination change from the perimeter toward their center.
For AIA~1600 intensity, the distributions beyond the 1-boundary (the original perimeter) closely resemble each other and become almost indistinguishable at the 3-boundary (i.e., 3 pixels, 1.8 arcseconds, from the perimeter).
Magnetic field distributions show large changes between the first few boundary layers but also become almost indistinguishable at 3 pixels from the perimeter.
The fact that the distributions of AIA~1600 intensity and magnetic field strength differ for the first few boundary layers also has implications for their mutual dependence.
The right column of Fig.~\ref{fig:boundaryrelations} shows that $I_\mathrm{AIA}-B_\mathrm{LOS}$ relation changes significantly for the first few boundary layers of the activity cluster.
On the activity cluster perimeter, the magnetic field is much weaker than expected from the median relation of all activity cluster pixels.
The $B_\mathrm{LOS}-I$ relation (the left column of Fig.~\ref{fig:boundaryrelations}) also changes somewhat between the first few layers, but their differences are much smaller.

\begin{figure}[!htbp]
\resizebox{\hsize}{!}{\includegraphics{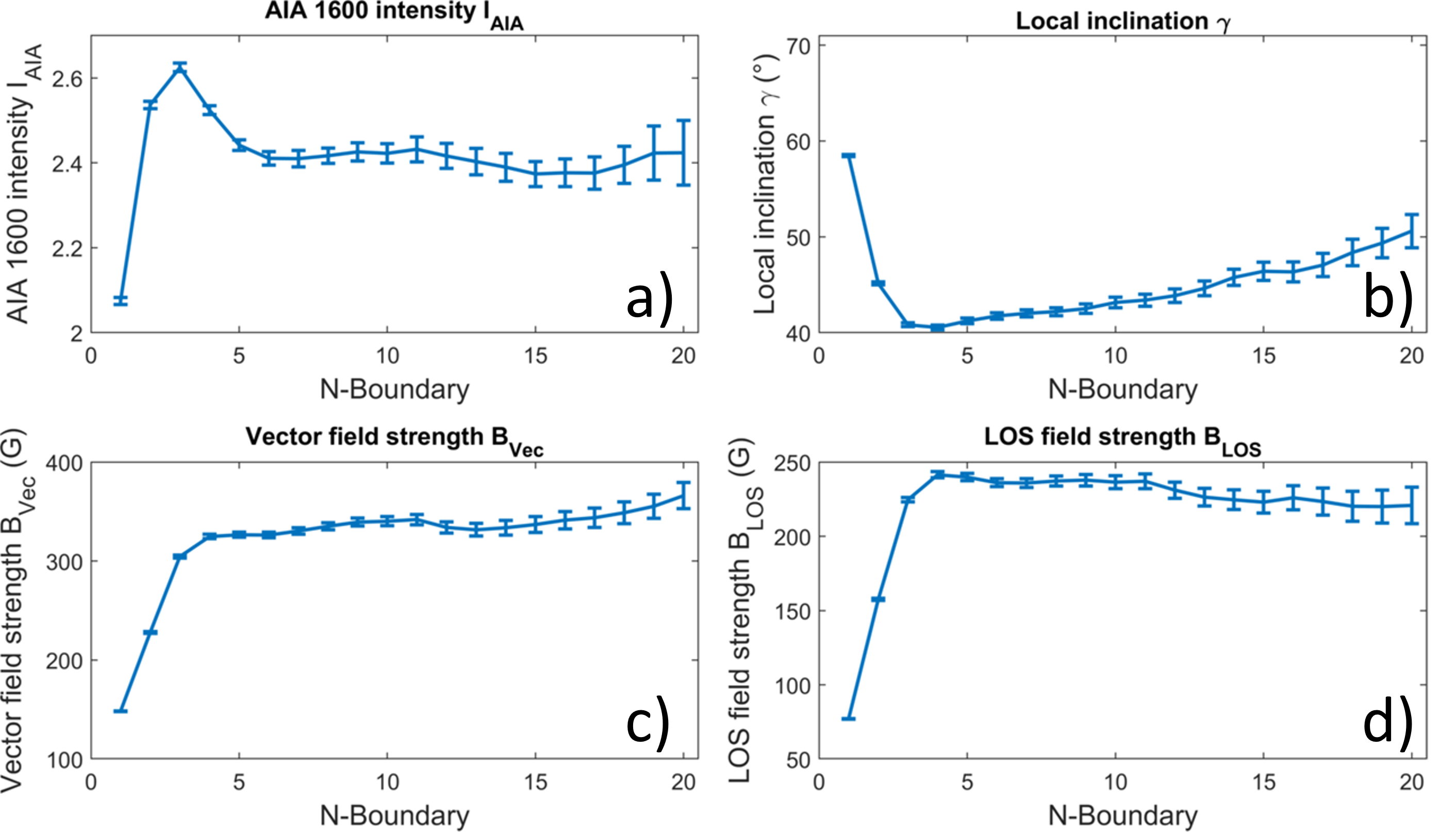}}
\caption{Boundary layer averages for A $>$ 1000 px activity clusters. (a) AIA 1600 intensity. (b) Local inclination. (c) Vector field strength. (d) LOS field strength. Error bars correspond to twice the standard error of the mean.
}\label{fig:BoundaryMeans}
\end{figure}

The above-discussed properties of the outermost boundary layers of the activity clusters reflect the change from the magnetically active region to the quiet Sun.
The average vector field strength and the average LOS field strength of the 1-boundary are 148 G and 77 G, respectively, and comparable to the thresholds that we used to separate the activity clusters from the quiet Sun.
Figure \ref{fig:BoundaryMeans} demonstrates how the mean values of AIA 1600 intensity, vector field strength, LOS field strength, and local inclination change during the transition from the original perimeter (the 1-boundary) to the inner N-boundaries up to N = 20.
Both magnetic field strengths (lower row of Fig.~\ref{fig:BoundaryMeans}) show only relatively small changes beyond N = 3.
Vector field strength more than doubles from 148 G to 325 G and LOS field strength more than triples from 76 G to 242 G from the 1-boundary to the 4-boundary.
Beyond N = 4, the LOS field strength stays about constant, while the vector field strength slightly increases simultaneously with inclination (panel b of Fig.~\ref{fig:BoundaryMeans}).
The AIA 1600 intensity (panel a of Fig.~\ref{fig:BoundaryMeans}) is also lowest at the 1-boundary, but the change to the roughly constant core level is smaller than for the two magnetic field strengths.
This is related to the fact that the AIA 1600 intensity at the 1-boundary is already more than twice the intensity of the quiet Sun.
Interestingly, the AIA 1600 intensity peaks at the 3-boundary, reaching both the minimum and maximum N-boundary intensities within these outermost pixels.
Opposite to the AIA emission, the local inclination decreases rapidly from its close to a maximum value at the 1-boundary to its minimum at the 4-boundary.
These changes underline the result that the outermost pixel layers separate from the core of the activity clusters quite distinctly.

\subsection{Relevance for other studies}
The result that both the magnetic field strength and local inclination affect the FUV emission around 1600~Å is important for detailed studies of the structure and dynamics of active regions.
The qualitative results on the mutual dependence of the parameters presented here will give important information, for example about the energetics of solar FUV emissions, active regions, and the chromosphere overall.
To our understanding, this is the first study where the detailed properties and mutual relations between FUV emission, magnetic field strength, and inclination have been studied as a function of the distance from the outer boundary of active regions and where the effective thickness of the boundary for these parameters has been determined quantitatively.

The results presented in this study are also interesting for long-term studies that use observations of solar emissions to study past solar activity.
For example, the Ca II K emission, which is closely related to AIA~1600 \citep{Rutten1999,Loukitcheva2009,Bose2018,Tahtinen2022} and for which there exists a century of observations, is used to reconstruct the photospheric magnetic field \citep{Pevtsov.etal2016,Chatzistergos2019,Virtanen2022}.
The reconstructed magnetic field can also be used to model solar irradiance over the past century \citep{Chatzistergos2021}.

\begin{figure}[!htbp]
\resizebox{\hsize}{!}{\includegraphics{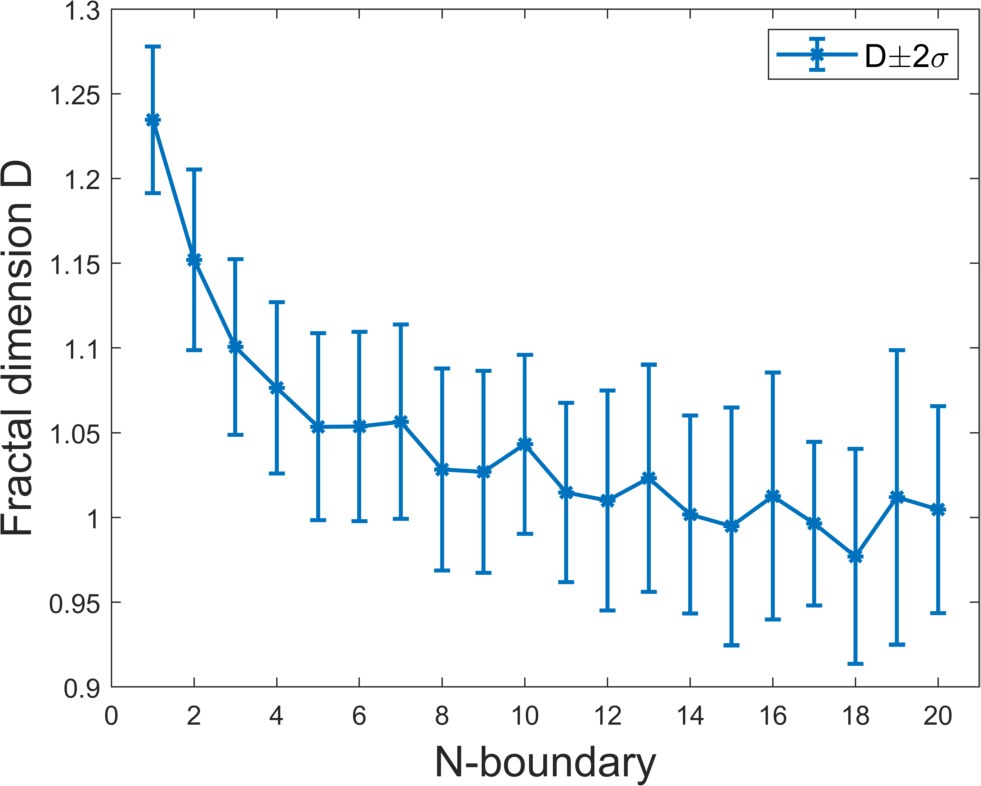}}
\caption{Average fractal dimension of the first 20 N-boundaries. The error is $\sigma = \sqrt{\sigma_\mu^2 + \sigma_D^2}$, where $\sigma_\mu$ is the standard error of the mean and $\sigma_D$ is the average standard error of the fractal dimension of individual activity clusters.
}\label{fig:BoundaryDimension}
\end{figure}

These historical long-term observations used a much lower spatial resolution than the modern high-resolution satellite observations used here.
This essentially complicates the direct comparison of our results with the historical observations.
With the AIA and HMI high-resolution observations and definition of the activity clusters used here, more than half (54.5\%) of all activity cluster pixels are within 3 pixels of the original perimeter for clusters larger than 1000 pixels.
This large percentage of the three first outer boundary pixels indicates that these outer layers of activity clusters have a fractal nature.
We calculated the average fractal dimension of activity clusters for each N-boundary up to N = 20 (see Fig.~\ref{fig:BoundaryDimension}).
One can see that the fractal dimension is largest and quite significantly larger than unity (about 1.23$\pm0.04$) for N = 1 and decreases rapidly until reaching values close to 1 (within $2\sigma$) from N = 5 onward.
This change from a large fractal dimension to values close to unity shows that the activity clusters have a complex outer boundary that quickly becomes more regular and less complex with increasing N.
We calculated the fractal dimension of each activity cluster (A $>$ 1000 px) using the box-counting method, which involves overlaying a grid of boxes of a linear size $l$ over the image and counting the number of boxes $N(l)$ that contain the cluster boundary (\citealp{Mandelbrot1982}, see also \citealp{McAteer2005} for solar-related studies).
This process is repeated for various box sizes.
The fractal dimension $D$ is determined from the power-law relationship between the box size and the number of boxes needed to cover the boundary $N(l) \propto l^{-D}$.

The fact that the first three boundaries with quite different relations between the AIA emission, magnetic field strength, and inclination for such a large fraction of all activity cluster pixels makes it difficult to estimate their contribution in low-resolution observations in which the boundary and core regions are averaged within a low-resolution element.
For example, the typical resolution of a historical, low-resolution synoptic map is one degree in longitude and 1/90 in sine of latitude, which at the equator translates to a pixel of the size $16.8''\times10.6''$, while the size of AIA and HMI pixels is $0.6''\times0.6''$ (i.e., the pixel size of a synoptic map is almost 500 times larger).
The effect of the outer boundary of active regions to the low-resolution observations can be studied reducing the resolution of AIA and HMI observations.
Such a study is underway but remains beyond the scope of this work.

\section{Conclusions}\label{sec:conclusions}
We have shown quantitatively how the inclination of the photospheric magnetic field affects FUV emission around 1600~Å.
In general, the AIA~1600 emission increases with increasing magnetic field strength and decreases with increasing inclination.
As in earlier studies of chromospheric emissions, we also find that the relation between AIA~1600 emission and the magnetic field strength is nonlinear, with AIA~1600 emissions saturating at a high magnetic field strength.
An important finding of this paper is that the peak intensity at which AIA~1600 emissions saturate depends on the cosine of the magnetic field inclination, with emissions from more horizontal regions saturating at a lower intensity.
We note that a similar cosine dependence has been found between the cutoff frequency of magnetoacoustic waves and the magnetic field inclination \citep{Bel1977,Rajaguru2019,Yelles2023}.
Our results also show that the activity clusters have a narrow boundary of less than~2 arcseconds, in which the average properties of AIA 1600 intensity, magnetic field strength, and local inclination, as well as their mutual relations, differ significantly from those in the inner layers.

The results presented in this paper are interesting for studies and models of the detailed structure and energetics of active regions, as well as for studies that aim to reconstruct past solar activity (magnetic fields or irradiance) based on chromospheric emissions.
We have studied the FUV emission around 1600~Å, but similar results are expected for studies that examine other parts of the continuum or various spectral lines.
For example, the Ca II K emission, which is often used to reconstruct past magnetic fields, resembles the AIA~1600 emission extremely closely both in terms of its dependence on the magnetic field strength and its spatial emission structure \citep{Tahtinen2022}.
We find that the activity cluster boundary has a complex fractal-like structure, which complicates a direct comparison of the results presented here with low-resolution observations.
Although we expect that the effect of an observed narrow activity cluster boundary will be significantly reduced in low-resolution observations, a detailed study is needed to examine the emission--magnetic field relation at different resolutions.

\begin{acknowledgements}
    I.T. and T.A. acknowledge the ﬁnancial support by the Research Council of Finland to the PROSPECT (project no. 321440). I.T. acknowledges the ﬁnancial support by the Finnish Academy of Science and Letters (Väisälä Fund) and the Jenny and Antti Wihuri Foundation.
    T. A. acknowledges the ﬁnancial support by the Research Council of Finland SOLEMIP (project no. 357249).
    K.M. acknowledges AURA/NSO for a visiting researcher grant.
    We thank the anonymous referee for the constructive comments that improved this article.
    The National Solar Observatory is operated by the Association of Universities for Research in Astronomy (AURA), Inc., under a cooperative agreement with the National Science Foundation.
    Authors acknowledge the international teams “Modeling Space Weather And Total Solar Irradiance Over The Past Century” (team \#475) and “Reconstructing Solar and Heliospheric Magnetic Field Evolution Over the Past Century” (team \#420) supported by the International Space Science Institute (ISSI), Bern, Switzerland and ISSI-Beijing, China.
    SDO data was downloaded from JSOC data facility courtesy of AIA and HMI science teams.
\end{acknowledgements}

\bibliography{bibliography} 

\begin{appendix}
    \section{LOS inclination}\label{sec:appendix}
\begin{figure} 
\resizebox{\hsize}{!}{\includegraphics{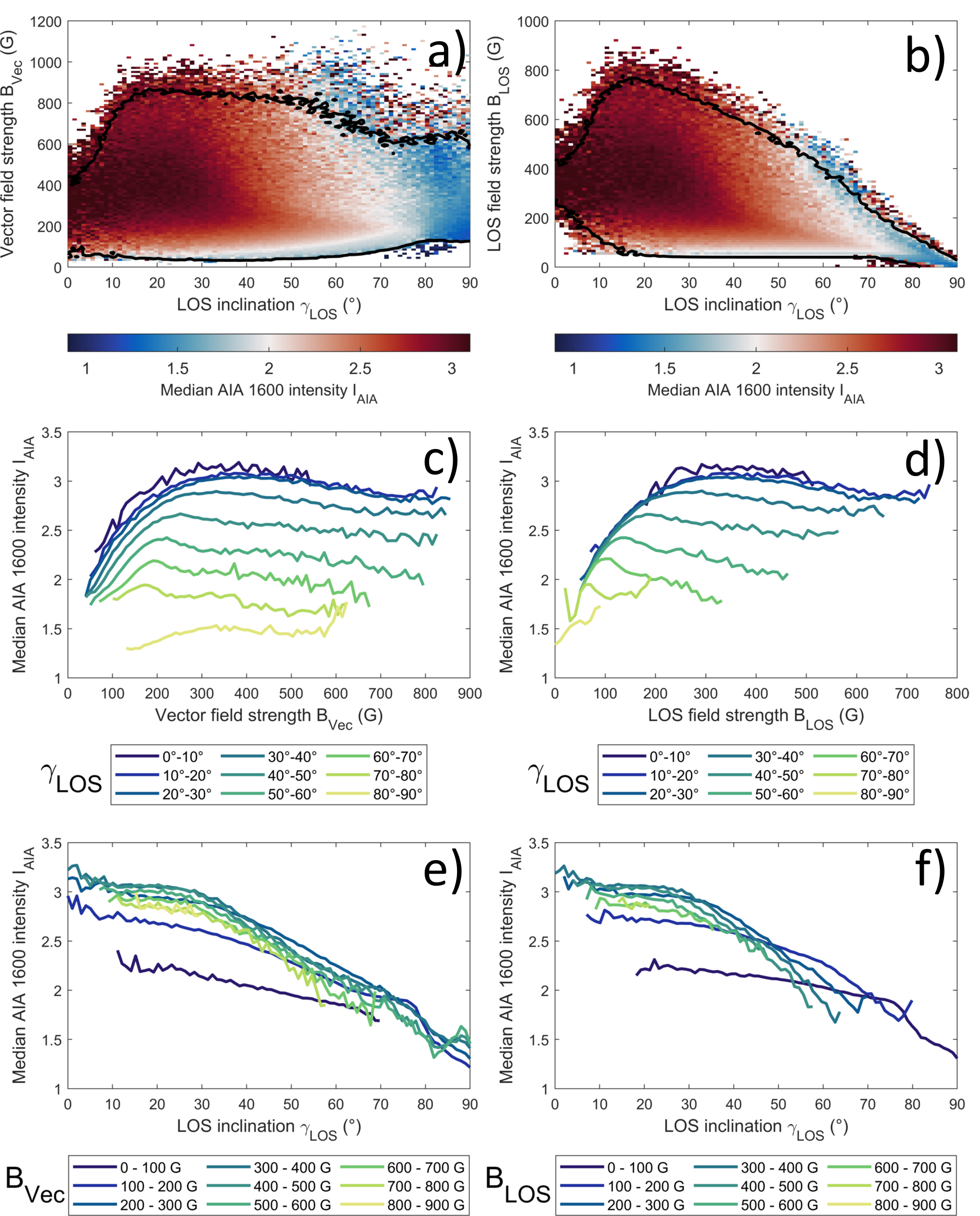}}
\caption{Median AIA~1600 intensity as a function of LOS inclination and magnetic field strength.
(a) Median AIA~1600 intensity as a function of LOS inclination and $B_\mathrm{Vec}$.
The bin size is $\mathrm{\Delta} \gamma_\mathrm{LOS}$~=~1\degree{} for LOS inclination and $\mathrm{\Delta} B_\mathrm{Vec}$~=~10~G for vector field strength.
The median AIA~1600 intensities are color-coded. White corresponds to the intensity level $I_\mathrm{AIA}~=~2$, i.e., twice the average level of the quiet Sun.
The black line shows the contour for bins with at least ten observations.
(b) Same as panel a, but for $B_\mathrm{LOS}$. The black line shows the contour for bins with at least ten observations.
}\label{fig:LOSInc}
\end{figure}

We performed the same analysis as presented in Sect.~\ref{sec:Inclination} using LOS inclination $\gamma_\mathrm{LOS}$ instead of local inclination $\gamma$.
The results are shown in Fig.~\ref{fig:LOSInc}, which is the same as Fig.~\ref{fig:AIA_vs_BandLOS_and_Inc} but shows the LOS inclination, $\gamma_\mathrm{LOS}$, instead of the local inclination, $\gamma$.
The results obtained using $\gamma_\mathrm{LOS}$ are in close agreement with the results obtained using $\gamma$, which confirms that possible problems that may result from the azimuthal disambiguation of the vector field have a negligible effect on our analysis.
The results are almost identical because the difference between $\gamma_\mathrm{LOS}$ and $\gamma$ is at most 12.95\degree{} within the selected 0.1$R_{\odot}$ distance from the disk center.

\end{appendix}

\end{document}